\pdfoutput=1
\documentclass[11pt]{iopart}

\usepackage{iopams}  
\expandafter\let\csname equation*\endcsname=\relax
\expandafter\let\csname endequation*\endcsname=\relax

\usepackage{physics}                            
\usepackage{bm}									
\usepackage{enumerate}							
\usepackage{amssymb}							
\usepackage[dvipsnames]{xcolor}					
\usepackage[colorlinks=true,allcolors=RoyalBlue]{hyperref}
\hypersetup{colorlinks=true}
\usepackage[capitalize]{cleveref}               
\usepackage{graphicx,color}
\usepackage{bbold}                              
\usepackage{cite}
\usepackage{tikz}
\usepackage[utf8]{inputenc}
\usepackage[T1]{fontenc}
\usepackage{braket}
\usepackage[normalem]{ulem} 
\usepackage{comment}

\makeatletter
\renewcommand{\thefootnote}{\arabic{footnote}}
\def\@fnsymbol#1{\arabic{#1}}
\renewcommand\@makefnmark{\hbox{\textsuperscript{\thefootnote}}}
\long\def\@makefntext#1{\noindent\hbox{\textsuperscript{\thefootnote}}#1} 
\makeatother
\usepackage{etoolbox}
\makeatletter
\newrobustcmd{\fixappendix}{%
  \patchcmd{\l@section}{1.5em}{7em}{}{}%
  \patchcmd{\l@subsection}{2.3em}{7em}{}{}%
}

\makeatother

\newcommand{\thetitle}{Level statistics in the fractal phase of generalized Rosenzweig--Porter models}

\newcommand{\n}{\nonumber}
\def \ss {_{\rho \rho}}

\def \ij {_{ij}}
\def \ee {\, \mathrm e}
\def \ii {\, \mathrm i}
\def \z {^{(0)}}
\def \o {^{(1)}}
\def \free {\kappa^{(\text{free})} } 
\def \cum {\kappa^{(\zeta)}} 

\def \etanew {\sigma}
\newcommand{\rhoemp}{\rho^{(\mathrm{e})}}
\def \newb {\textbf{M}}

\newcommand{\rev}[1]{#1}

\date{\today}

\begin{document}
\title[\thetitle]{\thetitle}
\author{Victor Delapalme$^{1}$, Leticia F.~Cugliandolo$^{2}$, Alexander K. Hartmann$^{3}$, Marco Tarzia$^{4}$, and Davide Venturelli$^{4}$}
\address{$^1$ Department of Mathematics, King’s College London, Strand, London, WC2R 2LS, UK}
\address{$^2$ Sorbonne Université, Laboratoire de Physique Théorique et Hautes Energies, CNRS-UMR 7589, 4 Place Jussieu, F-75252 Paris Cedex 05, France}
\address{$^3$ Institut f\"ur Physik, Universit\"at Oldenburg, D-26111 Oldenburg, Germany}
\address{$^4$ Sorbonne Université, Laboratoire de Physique Théorique de la Matière Condensée, CNRS-UMR 7600, 4 Place Jussieu, F-75252 Paris Cedex 05, France}

\begin{abstract}
The Rosenzweig--Porter (RP) random matrix ensemble has emerged as a minimal model for the integrability-to-chaos crossover in quantum many-body systems. 
Its phase diagram features a region with fractal eigenstates, exhibiting intermediate spectral and localization properties between the fully localized and fully delocalized regimes.
In this work, we explore several generalizations of the RP model and determine their level statistics at the scale of the Thouless energy $E_T$, which characterizes the crossover. 
Using tools from free probability theory and the replica method, we compute the full counting statistics in the limit of large system size, and show that it takes a simple, universal scaling form around $E_T$, shared across all variations of the model. 
We validate our analytical predictions using exact numerical diagonalization of large samples, and large-deviation algorithms that resolve the full counting statistics down to probabilities as low as $10^{-40}$.
We also contrast our predictions with measurements on the quantum random energy model, which is the simplest model displaying many-body localization. 
\end{abstract}

\vspace{10pt}
\begin{indented}
\item[]\today
\end{indented}

\tableofcontents

\markboth{\thetitle}{\thetitle}

\setcounter{footnote}{0} 

\section{Introduction}
\label{sec:introduction}

\subsection{\rev{Context}}

Understanding when and how isolated quantum systems evolve toward thermal equilibrium 
is a central problem in modern statistical mechanics, 
fueled by the experimental control of 
ultracold atomic gases, trapped ions, and other platforms in which the coherent dynamics of many-body quantum systems can
be probed with high precision.

Three kinds of isolated quantum systems can be broadly distinguished. 
Chaotic systems scramble quantum information and act as their own heat bath, 
erasing all local memory of initial conditions. Integrable systems possess an extensive set of conserved quantities constraining their dynamics, 
and retain information about all conserved charges. In many-body localized systems, strong disorder prevents 
the transport of information, and effectively freezes the system's memory of its initial state~\cite{Gornyi_2005,basko2006metal} (see Refs.~\cite{reviewMBL,reviewMBL2,reviewMBL3,reviewMBL4,reviewMBL5,sierant2024manybody,CuMu22} for recent reviews). 

The primary framework used to explain thermalization in chaotic systems 
is the Eigenstate Thermalization Hypothesis (ETH)~\cite{srednicki1994chaos,rigol2008thermalization}. 
ETH posits that in a chaotic quantum system
thermalization is encoded at the level of single eigenstates, which act as microcanonical ensembles themselves,
rather than emerging from dephasing over many of them. More precisely, 
ETH makes two related but distinct statements about the structure of chaotic eigenstates. 
First, it states that energy eigenstates are not localized in any physically natural basis, but instead spread ergodically over the entire accessible Hilbert space, with expansion coefficients that behave as pseudo-random Gaussian variables. 
Second, it proposes that the diagonal matrix elements of few-body observables
are smooth functions of the energy alone, agreeing with the microcanonical average at 
that energy, while the off-diagonal elements are random and exponentially small in the system size.
Therefore, any initial state with a sufficiently narrow energy window will relax to the microcanonical prediction for all local observables, since all eigenstates it overlaps with give essentially identical expectation values.
ETH is universally expected to apply to non-integrable, locally interacting many-body systems at finite 
temperature or high energy densities. Conversely, it breaks down in integrable systems and many-body localized phases.

The ETH structure finds a natural explanation in random matrix theory~\cite{Livan_2018,potters2020first,Mingo2017}: treating the eigenstates of a chaotic Hamiltonian as Haar-random vectors in Hilbert space --- as in the Gaussian Orthogonal and Unitary Ensembles (GOE and GUE) --- the statistical properties of both the expansion coefficients and the observable matrix elements follow directly, making random matrix theory the natural microscopic foundation of ETH.

\begin{figure}[t]
\begin{center}
\begin{tikzpicture}
\node[] at (-7.0,2.6) {\textcolor{darkgray}{\small Eigenvalue distribution}};
\node[] at (-7.0,1.5) {\textcolor{darkgray}{\small Level spacings}};
\node[] at (-7.0,0.75) {\textcolor{darkgray}{\small Eigenvectors}};
\draw[RoyalBlue, thick,->] (-5,2) -- (4,2);
\draw[gray, thick] (-1,1.9) -- (-1,2.1);
\draw[gray, thick] (0.5,1.9) -- (0.5,2.1);
\draw[black, dashed] (0.5,0.5) -- (0.5,1.5);
\draw[black, dashed] (-1,1.9) -- (-1,2.35);
\draw[black, dashed] (-1,2.75) -- (-1,3);
\draw[blue, dotted] (0.5,0.5) -- (0.5,1.5);
\draw[blue, dotted] (-1,0.5) -- (-1,2);
\node[] at (-3,2.6) {\textcolor{black}{$\rho_{\newb}$}};
\node[] at (1.5,2.6) {\textcolor{black}{$p_a$}};
\node[] at (-3,0.75) {\textcolor{gray}{\footnotesize fully extended}};
\node[] at (-0.25,0.75) {\textcolor{darkgray}{\footnotesize fractal}};
\node[] at (2,0.75) {\textcolor{black}{\footnotesize fully localized}};
\node[] at (-1.25,1.5) {\textcolor{black}{\footnotesize Wigner}};
\node[] at (2,1.5) {\textcolor{black}{\footnotesize Poisson}};
\node[] at (4,1.65) {\textcolor{black}{$\gamma$}};
\node[] at (-1,2.55) {\textcolor{NavyBlue}{ $1$}};
\node[] at (0.5,1.65) {\textcolor{NavyBlue}{$2$}};
\end{tikzpicture}
\end{center}
\caption{The phase diagram of the GRP model (see \cref{sec:model}).}
\label{fig:phase-diagram}
\end{figure}
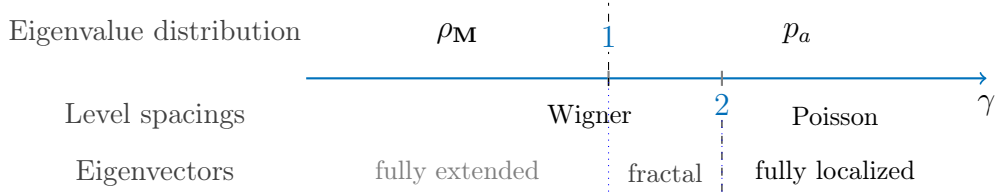

Now, understanding how ETH breaks down as one moves away from the GOE or GUE limit is a central question. 
In the quest for a unified framework describing thermalization in quantum systems, 
numerous random matrix ensembles were suggested as toy models to capture 
transitions from chaotic to non-chaotic regimes, with eigenvalue statistics serving as the primary diagnostic of such transitions~\cite{craps2026,Mirlin96,Fyodorov_2009,roy2018multifractality,wang2016phase,nosov2019correlation,duthie2022anomalous,kutlin2021emergent,motamarri2021localization,tang2022non,tarzia2022fully,Skvortsov_2022,Cai_2013,DeGottardi_2013,Liu_2015,Das_2022,Ahmed_2022,Lee_2022,Venturelli_2023,Das2023,LeDoussal2025,Baron2025,Suntajs2022,Kravtsov_2015,vonSoosten_2019,Facoetti_2016,Truong_2016,Bogomolny_2018,DeTomasi_2019,amini2017spread,pino2019ergodic,berkovits2020super,kravtsov2020localization,khaymovich2020fragile,monthus2017multifractality,biroli2021levy,buijsman2022circular,khaymovich2021dynamical,Sarkar_2023,Buijsman2024,cadez2024,jahnke2025,Ghosh_2025,Lunkin2025,safonova2025,swietek2026,Kutlin_2024}. 
Among the simplest and most paradigmatic of these is the Rosenzweig--Porter (RP) model~\cite{RP_1960}, 
originally introduced to describe complex atomic spectra, and since then the subject of extensive investigation~\cite{Kravtsov_2015,vonSoosten_2019,Facoetti_2016,Truong_2016,Bogomolny_2018,DeTomasi_2019,amini2017spread,pino2019ergodic,berkovits2020super,kravtsov2020localization,khaymovich2020fragile,monthus2017multifractality,biroli2021levy,buijsman2022circular,khaymovich2021dynamical,Sarkar_2023,Buijsman2024,cadez2024,jahnke2025,Ghosh_2025,Lunkin2025,safonova2025,swietek2026,Kutlin_2024} owing to the remarkably rich phenomenology that emerges from its apparent simplicity. 
The RP model is defined as the sum of two independent $N \times N$ random matrices, a diagonal random matrix  $\textbf{A}$ with independent and identically distributed (i.i.d.)~entries drawn from a distribution $p_a$, and a matrix $\textbf{B}$ belonging to the GOE:
\begin{equation}
    \textbf{H}_{\rm RP} = \textbf{A} + \nu N^{-\gamma/2}\, \textbf{B} \;,
    \label{eq:hamiltonian}
\end{equation}
with $\nu\sim \order{1}$.
Figure~\ref{fig:phase-diagram} illustrates the three 
distinct phases of the model: a fully delocalized/chaotic phase for $\gamma <1$, a fully localized phase for $\gamma > 2$, and an intermediate fractal phase for $1<\gamma <2$. 
Hence, by independently controlling the variance of the diagonal and off-diagonal random matrix entries through the parameter $\gamma$,
the RP model interpolates between a fully ergodic phase where GOE statistics and ETH hold, an intermediate phase where eigenstates are fractal
(see \cref{sec:fractal})
and ETH is violated in a subtle way, and a fully localized phase where ETH fails completely.

In this context, level statistics serve as a key probe of chaos in quantum systems. The simplest observable one can construct is the distribution of spacings between consecutive energy levels (Poisson vs Wigner--Dyson~\cite{potters2020first}); 
however, this approach suffers from two drawbacks. First, since the density of states is generally non-uniform across the spectrum, raw spacings reflect the global shape of the spectrum rather than the local chaotic or non-chaotic behavior of the system. This issue is conventionally addressed through the unfolding procedure, which normalizes the spectrum to enforce a uniform mean level spacing, or more conveniently by considering the ratios of consecutive spacings~\cite{Atas_2013}, defined as 
$r_n = \min(\delta_n, \delta_{n+1})/\max(\delta_n, \delta_{n+1})$, 
which are by construction independent of the spectral shape and require no unfolding. Second, spacing distributions only capture short-range correlations between neighboring levels, and are therefore blind to correlations extending over larger portions of the spectrum. 
A particularly elegant and informative approach that resolves both issues simultaneously is to study the full counting statistics of energy levels \rev{--- namely, the probability distribution of the number $ I_N[-E,E]$ of levels falling within the interval $[-E,E]$. In particular, the level compressibility~\cite{Mirlin_2000}, which is the ratio of the first two cumulants of $ I_N[-E,E]$, actually encodes the}
spectral correlations at all scales (see Section~\ref{sec:FCS} for precise definitions of these quantities).

Interestingly enough, in the intermediate phase of the 
RP
model, the level compressibility crosses over from RMT statistics at short scales (of the order of the spectral gap $\sim 1/N$), to Poisson statistics at large scales ($E\sim \mathcal{O}(1)$)~\cite{Venturelli_2023}. From a physical perspective, this crossover is related to the presence of mini-bands in the energy spectrum~\cite{Altshuler_2023} (this concept is recalled in Section~\ref{sec:model}). 
Crucially, the level compressibility in the crossover region 
\rev{turns out to assume a scaling form that is largely}
independent of the microscopic details of the model, up to a 
non-universal 
rescaling~\cite{Venturelli_2023,Delapalme2026,safonova2025,Kravtsov_2015}. In particular, it is insensitive to the choice of the distribution $p_a$ of the diagonal elements~\cite{Venturelli_2023}, to the presence of heavy tails in the distribution of the entries of $\mathbf{B}$~\cite{safonova2025}, to relaxing the assumption that the entries of $\mathbf{B}$ are independent and uncorrelated (at least to some extent)~\cite{Delapalme2026}, and to 
\rev{the fact that the Gaussian entries of $\mathbf{B}$ may be chosen complex (as in the GUE) rather than real~\cite{Kravtsov_2015}.}
It was 
already conjectured in~\cite{Venturelli_2023} that the specific choice of the off-diagonal elements (i.e.~the elements of $\textbf{B}$ in \cref{eq:hamiltonian}) would not affect the scaling function altogether, 
and that the latter may in fact be universal.

\subsection{\rev{Purpose of this work}}

In this paper, we address the universality conjecture put forward in~\cite{Venturelli_2023}\rev{, with the two-fold aim of extending its limits of validity, and of better understanding its scope}.
To this end, we \rev{first} generalize the definition of the RP model by allowing $\mathbf{B}$
to belong to one of two broad classes of random matrices: Wigner matrices 
and orthogonally invariant matrices
(the intersection of these two classes being the GOE). 
Wigner matrices, defined in Section~\ref{sec:wigner}, form a natural and well-studied ensemble whose defining feature is the statistical independence of their entries~\cite{Livan_2018,cipolloni2026rosenzweigportertypemodel}. Orthogonally invariant matrices, on the other hand, are the central objects of free probability theory~\cite{voiculescu1992free}, and have recently appeared in works on ETH and quantum chaos~\cite{vallini2025, camargo2025}. 
These are random matrices $\textbf{M}$ that are statistically invariant under rotations or, equivalently, 
whose probability distribution is rotationally invariant, i.e.~$P[ \textbf{M}  ] =  P [\textbf{OMO}^T]$ for any orthogonal matrix $\textbf{O}$.
Note that, with the exception of the Gaussian ensembles, all other members of this class have statistically dependent entries — Wishart-Laguerre matrices~\cite{Wishart1928, Delapalme2026} being a notable example. By contrast, this enlarged definition of the RP model does not include matrices whose entries have an infinite variance, as in the Lévy--RP model~\cite{biroli2021levy}.

Starting from such generalized definition of the RP model,
we compute the full counting statistics 
using 
the replica method, combined with tools from free probability theory for orthogonally invariant matrices. We show that, in the intermediate phase\footnote{\rev{As detailed in \cref{sec:model}, a proper normalization of the matrix elements can be introduced such that the interval $1<\gamma<2$ remains invariant across the variations of the model considered in this work.}} $1 < \gamma < 2$, 
both classes yield the same result up to 
finite-size corrections which vanish
in the large-$N$ limit. Furthermore, we derive the 
large-deviation function (LDF) of the full counting statistics, demonstrating that on the crossover scale, the cumulant generating function obeys a universal scaling form that is independent of the specific distributions of
the two random matrices in \cref{eq:hamiltonian}.
These analytical results are corroborated by
exact diagonalization, in combination with large-deviation approaches 
that allow
to access also the low-probability region of the counting statistics.

Driven by this observation, we address the question of which models violate the universal scaling form. 
A natural initial direction is to consider cases where the entries of $\textbf{B}$ are heavy tailed~\cite{monthus2017multifractality, biroli2021levy}; however, it was demonstrated in~\cite{safonova2025} that when $\textbf{B}$ is a L\'evy matrix, it yields the exact same scaling function for the 
level compressibility.\footnote{Indeed, the density-density correlation function computed in~\cite{safonova2025} can be directly related to the level compressibility --- see, e.g., Eq.~(58) in Ref.~\cite{Delapalme2026}.}
Alternatively, one could explore the scenario where the entries of $\textbf{A}$ are correlated~\cite{Altshuler_2023} --- a problem motivated by Hilbert space localization~\cite{Cohen16,Adler24,Jeyaretnam25} --- but we leave this avenue for future work. 
Instead, to assess the extent to which our results apply beyond abstract random matrix models, we investigate the case in which the assumption that the matrix $\mathbf{B}$ is dense (in the basis where $\mathbf{A}$ is diagonal) is relaxed, as is the case in realistic interacting many-body disordered systems represented on their Hilbert-space graph. 
As a testbed, we thus consider the Quantum Random Energy Model (QREM)~\cite{Faoro_2019,baldwin2018quantum,parolini2020multifractal,biroli2021out,kechedzhi2018efficient,Smelyanskiy_2020}, which is the simplest quantum mean-field spin glass. This choice is motivated by the fact that an analogy between the out-of-equilibrium phase diagram of the QREM and that of the RP model was previously established in Ref.~\cite{biroli2021out}, where it was shown that the latter captures the essential features of the former. In particular, the QREM exhibits a broad intermediate phase separating the fully ergodic and fully localized regimes, in which eigenstates are partially delocalized yet non-ergodic~\cite{Faoro_2019,baldwin2018quantum,parolini2020multifractal,biroli2021out,kechedzhi2018efficient,Smelyanskiy_2020}, closely resembling the intermediate phase of the RP model. It is therefore natural to ask whether, beyond the random matrix ensembles considered here, the QREM exhibits the same universal full counting statistics in this intermediate regime. 
Focusing on this specific regime, 
our numerical simulations reveal that the level compressibility converges to a distinct crossover function, which interpolates between RMT and Poisson statistics at a slower rate. 
We attribute this quantitative discrepancy to the fact that this spectral regime seems to host genuinely \textit{multi}fractal eigenstates rather than simply fractal, as is the case for the RP model. 
This discrepancy highlights the limitations of the standard RP paradigm and strongly motivates future investigations of the level 
compressibility within other models hosting multifractal eigenstates.

The rest of this manuscript is structured as follows. 
In Section~\ref{sec:model} we introduce the model
\rev{and the key quantities of interest, and we anticipate the main outcomes of our analysis.
In Section~\ref{sec:Path-integral},}
we 
frame within the replica formalism the calculation of
the full counting statistics for the generalization of the RP model to orthogonally invariant and Wigner matrices.
In Section~\ref{sec:scaling-function}, 
we obtain the cumulant generating function and the 
large-deviation function. 
We support our analytical predictions with numerical simulations in Section~\ref{sec:numerics}.
The numerical study of the level compressibility in the QREM is performed in Section~\ref{sec:QREM}. Finally, in Section~\ref{sec:conclusion} we summarize our results, and mention possible directions for future research.
\rev{Some important technical aspects are contained in the Appendix sections --- in particular, in~\ref{app:complex} we generalize our results  to Hermitian
matrices, in~\ref{app:E-O-1} we compute the cumulant generating function for energy windows of $\mathcal{O}(1)$, and in~\ref{app:FCS-independent} we compute the full counting statistics for independent energy levels as a benchmark.}

\section{The model}
\label{sec:model}

Our starting point is the $N\times N$ random matrix 
\begin{equation}
    \label{eq:RPdef}
    \textbf{H} = \textbf{A} + \sqrt{\etanew} \, 
    \newb
    \, ,
\end{equation}
which defines the generalized RP (GRP) model ~\cite{Venturelli_2023,Kravtsov_2015,vonSoosten_2019,Facoetti_2016,Truong_2016,Bogomolny_2018,DeTomasi_2019,amini2017spread,pino2019ergodic,berkovits2020super}.
The matrix $\textbf{A}$ is diagonal, $\textbf{A} \equiv \text{diag}(a_1, a_2, \dots, a_N)$, where $a_i$ are
i.i.d.~random 
variables sampled from the probability distribution $p_a(a)$ \rev{with variance of~${\mathcal O}(1)$}, and 
the matrix $\newb$ is drawn from the
GOE,
i.e.~its entries are real independent Gaussian random variables. 
In order for $\newb$ to have eigenvalues of $\mathcal{O}(1)$, we choose its entries to have a variance scaling as $\langle |m_{ij}|^2 \rangle \sim 1/N $ (in this sense the matrix $\newb$ differs from $\textbf{B}$ in \cref{eq:hamiltonian}, which in the original definition of the model has a spectrum with support scaling as $\sqrt{N}$). We also 
introduced the parameter 
\begin{equation}
\etanew = \nu^2 N^{1-\gamma} \, ,
\end{equation}
with $\nu$ and $\gamma$ of $\mathcal{O}(1)$. This choice of scaling allows one to preserve the phase diagram of the usual RP model~\cite{Venturelli_2023}, 
detailed below,
with the two transitions at $\gamma = 1$ and $ 2$. 

The RP model exhibits a rich phase diagram~\cite{Venturelli_2023,Kravtsov_2015, Facoetti_2016,khaymovich2021dynamical}, reproduced in Fig.~\ref{fig:phase-diagram}, 
governed by the variance of the off-diagonal matrix elements, relative to the mean level spacing. When the former is much smaller than the latter, $\gamma>2$, 
the system is in a localized phase, where the density of eigenvalues retains the discrete structure of the diagonal entries, and level spacings follow a Poisson distribution, reflecting the absence of level repulsion. 
As the variance of the off-diagonal matrix elements
increases beyond the mean level spacing, a first transition occurs toward an intermediate, partially delocalized phase, where eigenvalues begin to hybridize and the density of states broadens. A second transition takes place at
$\gamma=1$, beyond which the system enters a fully delocalized, ergodic phase: the density of eigenvalues converges to the Wigner semicircle law and level spacings follow the GOE Wigner--Dyson distribution, signaling the onset of strong level repulsion and quantum chaos.

An illuminating way to understand the phase diagram is to apply Fermi's Golden Rule to compute the transition rates induced by the perturbation $\sqrt{\etanew}\, \newb$ 
on the eigenstates of $\textbf{A}$~\cite{tarzia2020many,khaymovich2020fragile}, an approach that naturally introduces the Thouless energy. The
latter is a key energy scale which will play a central role in the subsequent analysis.
In this framework, the Thouless energy $E_T$ is defined as the width in the spectrum accessible in a time of $\mathcal{O}(1)$ by a particle undergoing quantum evolution induced by the Hamiltonian~\eqref{eq:hamiltonian} or,
equivalently, the width of energies over which  
states are hybridized by the perturbation:
\begin{equation}
    \label{eq:Thouless_energy}
    E_T \sim \sum_{j(\neq i )} \langle| h_{ij}|^2 \rangle = \etanew \sum_{j(\neq i )} \langle |m_{ij}|^2 \rangle \sim \etanew \sim N^{1-\gamma} 
    \; .
\end{equation} 
For $\gamma < 1$, $E_T \gg 1$ and the Thouless window coincides with the whole spectrum, making all levels reachable and eigenvectors delocalized. On the contrary, for $\gamma > 2$, $E_T \ll \delta _N  \sim N^{-1}$, where $\delta _N $ is the average spectral gap, meaning that the Thouless window is not wide enough to hybridize multiple states, thus making the energy levels 
effectively
independent, and eigenvectors fully localized.
Finally, for $1< \gamma < 2$, one has $ \delta _N \sim N^{-1} \ll E_T \ll 1$, meaning that the Thouless window contains a sub-extensive part of the spectrum. Only a fraction of levels are then reachable, and
eigenvectors are fractal, with fractal dimension $D_{\gamma} = 2-\gamma$ (see the next Section for a definition of fractality). Eigenvalues separated by more than the Thouless energy are effectively decoupled, while those within this window become correlated --- giving rise to the concept of mini-bands in the spectrum~\cite{Altshuler_2023}.

\subsection{Fractal properties of eigenstates}
\label{sec:fractal}

A useful way to characterize non-ergodic wave-functions is via their fractal dimensions $D_q$.
These can be estimated from the scaling with the system size $N$ of the moments of the wave-function coefficients --- referred to as the generalized inverse participation ratios (IPRs):
\begin{equation}
\label{eq:ipr}
I_{ q} = \sum_i |\langle i|\psi\rangle|^{2{q}} \propto N^{D_{ q} (1-{q})}
\; .
\end{equation}
The fractal dimensions take values in the range $0\leq D_q\leq 1$, where $D_q=0$ and $D_q=1$ correspond to fully localized and delocalized eigenstates, respectively. 

In the intermediate phase ($1< \gamma <2$) of the GRP model, $D_q=D_\gamma=2-\gamma<1$ becomes independent of $q$ (for $q>1/2$); hence, the corresponding eigenstates are fractal. Multifractal eigenstates with 
$q$-dependent fractal dimensions $D_q$ are known to exist at the critical point of Anderson localization~\cite{Mirlin_2000}. They have also been found 
numerically as the eigenvectors of
the weighted adjacency matrices of random Erd\"os-R\'enyi graphs~\cite{cugliandolo2024multifractal}.

\subsection{Generalizations considered in this work}
\label{sec:generalizations}

In this work, we consider the following two extensions of the standard RP model.
\begin{enumerate}[(i)]
    \item In \cref{sec:free}, we 
    take $\newb$ to be a random matrix drawn from a rotationally invariant  probability distribution. Equivalently, for matrices with real entries, this means that we can express $\newb = \textbf{OXO}^T$, where $\textbf{X}$ is a diagonal matrix and $\textbf{O}$ is a Haar-distributed orthogonal matrix. 
   We remark that, in this setting, the entries of $\newb$
   are generically not independent random variables. 
    Examples include the Gaussian ensembles, the Wishart--Laguerre ensemble, $\beta$--ensembles, or more generally, ensembles whose joint probability density function is of the form $\rho [M] \propto \ee^{-\Tr V(M)}$, with $V(x)$ a generic (analytic) function such that $\ee^{-\Tr V(M)}$ is normalizable~\cite{Livan_2018,potters2020first,Mingo2017}.
    \rev{Although we will focus for simplicity on real matrices, in fact generalizing our construction to complex Hermitian matrices only requires replacing orthogonal by unitary matrices, as we show in \ref{app:complex}.}

    \item In \cref{sec:wigner}, we consider the case in which $\newb$ is a Wigner matrix, constructed as follows. 
Let $(\zeta\ij)_{1\leq i, j\leq N}$ be i.i.d.~random variables distributed according to $p_\zeta(x)$. Then, $\newb$ is the real symmetric matrix with entries
\begin{equation}
    m_{ij} = (\zeta\ij+\zeta_{ji})/\sqrt{2N}.
    \label{eq:def_entries}
\end{equation} 
\end{enumerate}

In both cases, a useful probe of the local level statistics is the \textit{full counting statistics} that we introduce in the next Section.

\subsection{\rev{Summary of the main results of this work}}
\label{sec:FCS}

For a given realization of the random matrix $\textbf{H}$, we call $\lambda_i$ its eigenvalues, and let
\begin{equation}
    \rhoemp(\lambda) = \frac{1}{N} \sum_{i=1}^N \delta(\lambda-\lambda_i)
    \label{eq:emp_density}
\end{equation}
denote the ``empirical'' eigenvalue density. The quantity
\begin{equation}
    I_N[\alpha,\beta] \equiv N \int_{\alpha}^{\beta} \dd{\lambda} \; \rhoemp(\lambda)  
    = \sum_{i=1}^N  \; 
    [\Theta(\beta-\lambda_i) - \Theta(\alpha-\lambda_i)] 
    \label{eq:levels_number}
\end{equation}
then
denotes the number of eigenvalues $\lambda_i$ lying in the interval $ [\alpha,\beta]\subseteq \mathbb{R}$  (where $\Theta(x)$ is the Heaviside distribution).
Since
$I_N[\alpha,\beta]$ is in turn a random variable, 
in the following we set out to compute its cumulant generating function (CGF) in the limit of large $N$,
\begin{equation}
    \mathcal{F}_{[\alpha,\beta]}(s) 
    =\ln \langle \mathrm \ee^{-sI_N[\alpha, \beta]}\rangle.
    \label{eq:CGF-def}
\end{equation}
The expansion of this expression in powers of $s$ formally yields the scaled cumulants
\begin{equation}
    \frac{\kappa_j[\alpha,\beta]}{N} = (-1)^{j} \eval{\partial_s^j \mathcal{F}_{[\alpha,\beta]}(s)}_{s=0} \, .
    \label{eq:cumulants-general}
\end{equation}
In particular, upon choosing a symmetric interval $[\alpha,\beta]=[-E,E]$, the ratio of the first two cumulants defines the \textit{level compressibility} 
\begin{equation}
\label{eq:levelcomp-first}
 \chi(E) = \kappa_2(E)/\kappa_1(E)  \, .
\end{equation}
\rev{As recalled in the Introduction, when inspected over intervals $E\sim E_T$ of the order of the Thouless energy~\eqref{eq:Thouless_energy}, this quantity turns out to assume a simple scaling form $\chi(E) = \chi_T(E/E_T)$ (where $\chi_T(y)$ is given in Eq.~\eqref{eq:chi_universal} below) that is largely independent of the details of the matrices $\textbf{A}$ and $\textbf{M}$ in \cref{eq:RPdef}, and which was thus conjectured to be universal in Ref.~\cite{Venturelli_2023}. In this work, we take several steps forward:
\begin{enumerate}[(i)]
    \item Using the replica method delineated in \cref{sec:Path-integral}, we compute the full counting statistics~\eqref{eq:CGF-def} up to leading order for large system size $N$, in the fractal phase $1<\gamma <2$ of the model, and for both classes of random matrices introduced in Section~\ref{sec:generalizations}. We report this result in \cref{eq:CGF-E-O-1,eq:r-theta} below.
    \item For energies $E\sim E_T$, we show that not only the level compressibility~\eqref{eq:levelcomp-first}, but actually the whole cumulant generating function~\eqref{eq:CGF-def} assumes a simple scaling form (see \cref{eq:scaling-function-def,eq_scaling-function-G} below), depending only on the ratio $E/E_T$, and universal across all variations of the model described above. 
    \item We support our conclusions numerically by measuring both the level compressibility (see c.f.~\cref{fig:LC_numerics}) and the large-deviation function (see c.f.~\cref{fig:counting:rare}), the latter using an \textit{ad-hoc} algorithm able to resolve probabilities down to $10^{-40}$.
    \item To test the limits of validity of the putatively universal level compressibility, we measure this quantity on the Quantum Random Energy Model (QREM), which is one of the simplest interacting models displaying many-body localization. This model is clearly not part of the classes of random matrices covered by our analytical calculation, and yet its level compressibility turns out to display interesting collapse properties as a function of $E/E_T$ (see c.f.~\cref{fig:chiQREM}), which we discuss in \cref{sec:discussion-qrem}.
\end{enumerate}
}

\section{Replica path-integral formulation}
\label{sec:Path-integral}

The aim of this Section is to express the CGF in \cref{eq:CGF-def} in the form of a path integral over some auxiliary continuous fields, within the replica method.
\rev{Starting from the seminal work~\cite{Edwards_1976}, this method has been particularly fruitful, as it allows to treat random matrices on the same footing as classical disordered spin systems, and thus to leverage on the large body of literature related to them~\cite{Mezard_1987}.}
In this framework, the leading-order contribution to the CGF in the large-$N$ limit can be accessed via a saddle-point evaluation, which we later detail in \cref{sec:scaling-function}.

To this end, we first \rev{focus on the definition of $ I_N[\alpha,\beta]$ in \cref{eq:levels_number}, and we}
recall that the Heaviside function can be represented in terms of the discontinuity of the complex logarithm, 
\begin{equation}
    \Theta(-x) = \frac{1}{2\pi \ii}\lim_{\varepsilon \to 0^+} \left[ \ln(x+\ii\varepsilon) - \ln(x-\ii\varepsilon) \right] \, ,
\end{equation}
so that we can interpret
\begin{equation}
    \sum_{i=1}^N \Theta(\alpha-\lambda_i) = \frac{1}{2\pi \ii}\lim_{\varepsilon \to 0^+} \left[ \ln \det ( \textbf{H}- \alpha_\varepsilon 1 ) - \ln\det ( \textbf{H}- \alpha_\varepsilon^* 1 ) \right] \, ,
\end{equation}
where we denoted $\alpha_\varepsilon=\alpha -\mathrm i \varepsilon$ (the meaning of $\beta_\varepsilon$ below is analogous).
Next, we introduce the auxiliary partition function
\begin{equation}
    \mathcal{Z}(z) = \int_{\mathbb{R}^N} \! \mathrm{d}\textbf{r} \; \ee^{\pm\frac{{\rm i}}{2} \textbf{r}^{T}(z\textbf{I} - \textbf{H}) \,\textbf{r} }
    \, ,
\end{equation} 
with the $\pm$ sign corresponding to $\Im z > 0$ and  $\Im z < 0$,  respectively. This allows us to
rewrite $I_N[\alpha,\beta]$ as
\begin{equation}
    I_N[\alpha,\beta] = -\frac{1}{\pi {\rm i}}
    \lim_{\varepsilon \to 0^+} \ln \left[\frac{\mathcal{Z}(\beta-{\rm i}\varepsilon)\mathcal{Z}(\alpha+{\rm i}\varepsilon) }{\mathcal{Z}(\beta+{\rm i}\varepsilon)\mathcal{Z}(\alpha-{\rm i}\varepsilon)} \right].
\end{equation} 
 Assuming that the order of the limit $\varepsilon \to 0^+$ and the logarithm can be exchanged, one can 
 express the CGF of $I_N[\alpha,\beta]$ \rev{in \cref{eq:CGF-def}} as
\begin{equation}
    \mathcal{F}_{[\alpha,\beta]}(s) 
    = \frac{1}{N} \lim_{\varepsilon \to 0^+} \; \ln \; \langle \left[ \mathcal{Z}(\beta_\varepsilon^*)\mathcal{Z}(\alpha_\varepsilon)\right]^{{\rm i}s/\pi}  \left[\mathcal{Z}(\beta_\varepsilon) \mathcal{Z}(\alpha_\varepsilon^*)\right]^{-{\rm i}s/\pi}   \rangle 
    \; .
    \label{eq:CGF-replica}
\end{equation}
\rev{This quantity}
can now be evaluated within the replica method by first determining
\begin{equation}
    \label{eq:Q_n_pm}
    Q_{[\alpha, \beta]}(n_{\pm}) = \langle \left[ \mathcal{Z}(\beta_\varepsilon^*)\mathcal{Z}(\alpha_\varepsilon)\right]^{n_+}  \left[\mathcal{Z}(\beta_\varepsilon) \mathcal{Z}(\alpha_\varepsilon^*)\right]^{n_-}   \rangle 
\end{equation}
for integer $n_\pm$, 
from which \rev{the CGF~\eqref{eq:CGF-replica}} can then be obtained through analytic continuation as
\begin{equation}
    \label{eq:CGF_replica}
    \mathcal{F}_{[\alpha,\beta]}(s) =  
    \frac{1}{N}  \lim_{\varepsilon \to 0^+} \; \ln \; \lim_{n_{\pm} \to \pm {\rm i} s/\pi} \; Q_{[\alpha, \beta]}(n_{\pm})
    \, . 
\end{equation}


In most cases, computing $Q_{[\alpha, \beta]}(n_{\pm})$ exactly is unfeasible. 
In the \rev{following,}
we thus reformulate the problem within a path-integral framework, \rev{separately for the two classes of random matrices introduced in Section~\ref{sec:generalizations}}, which enables controlled approximations in the large-$N$ limit~\cite{Metz_2016,Metz_2017,Venturelli_2023}. We will mostly focus on intervals $[\alpha,\beta]=[-E_T,E_T]$ of the scale of the Thouless energy~\eqref{eq:Thouless_energy}, and we will work at \rev{$\order{\etanew}=\order{N^{1-\gamma}}$}.

\subsection{Orthogonally invariant matrices}
\label{sec:free}

Let us begin by considering the case in which the matrix $\newb$ in \cref{eq:RPdef} is any random matrix with a rotationally invariant statistical weight (see the definition in Section~\ref{sec:model}). The matrices $\textbf{A}$ and $\textbf{M}$ are then said to be mutually free, and the problem can be addressed using tools from free probability, as we do in the following.

In order to compute $Q_{[\alpha, \beta]}(n_{\pm})$ defined in Eq.~\eqref{eq:Q_n_pm}, one has to average over the realizations of the random entries of $\textbf{A}$ and $\newb$. This is conveniently achieved by first introducing the block matrices
\begin{equation}
    \label{eq:def_lambda_L}
        \underline{\Lambda} = \mqty(\dmat{
    \alpha_\varepsilon 1_{n_+} \!\!\!\!,
      -\beta_\varepsilon^*  1_{n_+} \!\!\!\!,
     \beta_\varepsilon 1_{n_-} \!\!\!\!, -\alpha_\varepsilon^*  1_{n_-} }),
    \qquad
    \underline{L} = \mqty(\dmat{
     1_{n_+}\!\!\!\!, - 1_{n_+} \!\!\!\!, 1_{n_-} \!\!\!\!, - 1_{n_-} }),
\end{equation} 
where $1_{n_\pm}$ is the identity in $n_\pm$ dimensions,
so as to express
\begin{align}
    \label{eq:step_1}
    &Q_{[\alpha, \beta]}(n_{\pm}) = \Biggl \langle  \int \prod_{i=1}^N \prod_{\rho=1}^n \mathrm{d}r_{i \rho} \, \exp \left [ - \frac{{\rm i}}{2} \sum_{i,j = 1}^N \sum_{\rho = 1}^n  r_{i \rho} \left( \lambda_{\rho \rho} \delta_{ij} - \ell_{\rho \rho} h_{ij} \right ) r_{j \rho}  \right] \Biggr \rangle_{\textbf{A},\newb}  \nonumber \\
    & = \Biggl \langle  \int \prod_{i=1}^N \mathrm{d}\Vec{r}_i \, \exp \left [ - \frac{{\rm i}}{2} \sum_{i = 1}^N  \Vec{r}_i^{\; T} \left( \underline \Lambda  - a_i \underline L \right ) \Vec{r}_i - \frac{i\sqrt{\etanew}}{2} \sum_{i,j = 1}^N  m_{ij}  \Vec{r}_i^{\; T} \underline{L} \Vec{r}_j\right] \Biggr \rangle_{\textbf{A},\newb} \nonumber \\
    & =   \int \prod_{i=1}^N \mathrm{d}\Vec{r}_i \, \ee^{-\frac{{\rm i}}{2} \sum_i \Vec{r}_i^{\; T} \underline \Lambda \Vec{r}_i} \times\Biggl \langle \exp \left ( \frac{{\rm i}}{2} \sum_{i = 1}^N    a_i \Vec{r}_i^{\; T}  \underline L  \Vec{r}_i  \right ) \Biggr \rangle_{\textbf{A}}  \Biggl \langle \exp \left ( \frac{{\rm i}\sqrt{\etanew}}{2} \sum_{i,j = 1}^N  m_{ij}  \Vec{r}_i^{\; T} \underline{L} \Vec{r}_j \right) \Biggr \rangle_{\newb} \nonumber \\
    & =   \int \prod_{i=1}^N \mathrm{d}\Vec{r}_i \, \ee^{-\frac{{\rm i}}{2} \sum_i \Vec{r}_i^{\; T} \underline \Lambda \Vec{r}_i} \times \prod_{i=1}^N \psi_a \left( -\frac{1}{2} \Vec{r}_i^{\; T} \underline \Lambda \Vec{r}_i \right ) \times \Biggl \langle \exp \left ( \frac{N}{2} \Tr  \textbf{MT} \right) \Biggr \rangle_{\newb} ,
\end{align} 
where we defined  the $N\times N$ matrix
\begin{equation}
\textbf{T} \equiv \frac{{\rm i} \sqrt{\etanew}}{N} \sum_{\rho=1}^n \ell_{\rho \rho} \textbf{r}_\rho \textbf{r}_\rho^T 
\; , 
\label{eq:T-def}
\end{equation} 
and 
\begin{equation}
    \psi_a(x) = \int \mathrm{d}a \; p_a (a) \ee^{-{\rm i} ax}
\end{equation}
is the characteristic function of $p_a$. Note that the vectors $\vec r_i \in \mathbb R^n$, 
$i=1,\dots,N$, have components $r_{i \rho}\in \mathbb{R}$, with $n=2(n_++n_-)$,
while the vectors $ \textbf{r}_\rho \in \mathbb R^N$, $\rho=1,\dots, n$, have components $r_{\rho i}\in \mathbb{R}$. We distinguish them using arrow and boldface, respectively.
Similarly, we distinguish the matrices with size proportional to replica indices
from the ones of linear size $N$ using underline plain and bold symbols, respectively. 
Finally, we use lower-case letters to denote the elements of the matrix with the corresponding upper-case name.

To make progress, we now recall the definition and some properties of the Harish--Chandra--Itzykson--Zuber (HCIZ) 
integral (see for instance Eqs.~(10.23) and~(10.45) in Ref.~\cite{potters2020first}, or App.~D in Ref.~\cite{vallini2025}).
For a given diagonal matrix $\textbf{X}$, we define 
\begin{equation}
    \label{eq:HCIZ}
    I ( \textbf{X}, \textbf{Y}) \equiv \Biggl \langle 
    \ee^{
    \frac{N}{2} \Tr 
    [\textbf{OXO}^T \textbf{ Y} ]
    }
    \Biggr \rangle_{\textbf{O}},
\end{equation} 
where $\langle \dots \rangle_{\textbf{O}}$ is the average\footnote{This construction works analogously for unitary matrices, upon inserting an additional factor of 2 in the exponential in \cref{eq:Low_rank_HCIZ}. \rev{Further details about the generalization to Hermitian matrices are given in \ref{app:complex}, while we focus here for clarity on the real (orthogonal) case.}\label{foot:unitary}} over the Haar measure of $O(N) $. 
In the cases in which $\textbf{Y} $ is of rank $\ll N$, and for large $N$, one obtains
\begin{equation}
    \label{eq:Low_rank_HCIZ}
    I ( \textbf{X}, \textbf{Y}) \approx \exp \left ( \frac{N}{2} \Tr \mathcal{H}_{\textbf{X}}(\textbf{Y}) \right ),
\end{equation} 
where $\mathcal{H}_{\textbf{X}} (t) \equiv \int_0^t \mathrm{d}s \, \mathcal{R}_{\textbf{X}}(s) $ is the anti-derivative of the R-transform associated to $\textbf{X}$. The latter can in turn be defined as the function satisfying
\begin{equation}
    \label{eq:def_R_transform}
    \mathcal{G}_{\textbf{X}}(z) = \frac{1}{z-\mathcal{R}_{\textbf{X}}(\mathcal{G}_{\textbf{X}}(z))}
    \; , 
\end{equation} 
where $\mathcal{G}_{\textbf{X}}(z) \equiv \Tr(\textbf{X}- z 1 )^{-1}$ is the resolvent associated to $\textbf{X}$. 
In Eq.~(\ref{eq:Low_rank_HCIZ}), the notation $ \mathcal{H}_{\textbf{X}} (\textbf{Y})$ denotes the matrix defined 
through the series expansion
\begin{equation}
    \label{eq:H_series}
    \mathcal{H}_{\textbf{X}}(z) = \sum_{j=1}^{+\infty} \frac{\free_j}{j} z^j ,
\end{equation} 
where $\free_j$ are the so-called free cumulants (see e.g.~Chapter~12 in Ref.~\cite{potters2020first}).

In our case, since $\newb = \textbf{OXO}^T$ is an orthogonally invariant matrix, $\textbf{O}$ is Haar distributed. Thus, the average over $\newb$ in \eqref{eq:step_1} is equivalent to averaging over the Haar measure of the $O(N)$ group and over the joint distribution of the diagonal matrix $\textbf{X}$, thus making this average an HCIZ integral. 
Furthermore, $\textbf{T}$ is a sum of $n$ rank-one matrices (see \cref{eq:T-def}), and it is thus at most of rank $n \ll N$. Therefore, we can use the Low-Rank HCIZ integral~\eqref{eq:Low_rank_HCIZ} to get
\begin{equation}
    \Biggl \langle \exp \left ( \frac{N}{2} \Tr  \textbf{MT} \right) \Biggr \rangle_{\newb} \approx 
     \Biggl \langle \exp \left ( \frac{N}{2} \Tr \mathcal{H}_{\textbf{X}}(\textbf{T}) \right ) \Biggr \rangle_{\textbf{X}}
     \; . 
\end{equation}
Note that $\mathcal{H}_{\textbf{X}}$ is indirectly a function of the resolvent $\mathcal{G}_{\textbf{X}}$. Now, a given realization of the random matrix $\textbf{X}$ (say $\textbf{X}_0$) formally produces an ``empirical'' resolvent $\mathcal{G}_{\textbf{X}}^{(e)}(z) \equiv \Tr(\textbf{X}_0- z \textbf{1} )^{-1}$; however, we expect that the latter will self-average when $N\to \infty$, so that $\mathcal{G}_{\textbf{X}}^{(e)}(z) = \mathcal{G}_{\textbf{X}}(z) + \mathcal{O}(N^{-1}) $. 
Therefore, up to corrections of order $N^{-1} $, we can safely omit the explicit average over $\textbf{X}$, to obtain
\begin{equation}
    \Biggl \langle \exp \left ( \frac{N}{2} \Tr \mathcal{H}_{\textbf{X}}(\textbf{T}) \right ) \Biggr \rangle_{\textbf{X}} \approx \exp \left ( \frac{N}{2} \Tr \mathcal{H}_{\textbf{X}}(\textbf{T})  \right )
    \; . 
\end{equation} 
Combining all these results, Eq.~\eqref{eq:step_1} becomes
\begin{align}
    \label{eq:step_2}
    Q_{[\alpha, \beta]}(n_{\pm}) \approx \int \prod_{i=1}^N \mathrm{d}\Vec{r}_i \, \ee^{-\frac{{\rm i}}{2} \sum_i \Vec{r}_i^{\; T} \underline \Lambda \Vec{r}_i} \times 
    \psi_a \left( -\frac{1}{2} \Vec{r}_i^{\; T} \underline \Lambda \Vec{r}_i \right ) \times 
    \ee^{
    \frac{N}{2} \Tr \mathcal{H}_{\textbf{X}}  \left ( \textbf{T}  \right ) 
    }
    \; . 
\end{align}
We now introduce the normalized density
\begin{equation}
    \label{eq:mu-density}
    \mu (\vec{y}) = \frac{1}{N} \sum_{i = 1}^{N} \prod_{\rho = 1}^{n} \delta(y_{\rho} - r_{i \rho}) = \frac{1}{N} \sum_{i = 1}^{N}  \delta(\vec{y} -\vec{r}_i) \, ,
\end{equation}
which can be inserted into~\eqref{eq:step_2} using the functional integral representation of the identity
\begin{equation}
    1 \propto \int \mathcal{D}\mu \, \mathcal{D}\tilde \mu \, \exp{-\ii \int \dd{\vec{y}} \tilde \mu(\vec{y}) \left[ N\mu(\vec{y}) - \sum_{i=1}^N \prod_{\rho=1}^n \delta (y_\rho-r_{i\rho}) \right] } \, ,
    \label{eq:identity}
\end{equation}
where $\tilde \mu (\vec y)$ is an auxiliary field.
Upon introducing the $N \times n $ matrix $\textbf{R}$, with entries  $r_{i \rho}$, we can then rewrite the matrix $ \textbf{T}$ in Eq.~\eqref{eq:T-def} as
\begin{equation}
    \textbf{T} = \frac{\rm i \sqrt{\etanew}}{N} \textbf{R} \underline L \textbf{R}^T
    \, ,
\end{equation}
where $\underline L$ was given in \cref{eq:def_lambda_L}.
The third factor on the right-hand side of Eq.~\eqref{eq:step_2} can be rewritten using the series expansion of $ \mathcal{H}_{\textbf{X}}  $ given in Eq.~\eqref{eq:H_series} and the cyclic property of the trace. The expression in the 
exponential then becomes
\begin{align} 
    &\frac{1}{2} \Tr \mathcal{H}_{\textbf{X}}   \left (  \textbf{T} \right) = \frac{1}{2} \Tr \mathcal{H}_{\textbf{X}} \left ( \ \frac{\rm i \sqrt{\etanew}}{N} \textbf{R} \underline L \textbf{R}^T  \right )  = \frac{1}{2} \sum_{j=1}^{+\infty} \frac{\free_j}{j} \Tr \left ( \ \frac{\rm i \sqrt{\etanew}}{N} \textbf{R} \underline L \textbf{R}^T  \right ) ^j   \\
    & \quad = \frac{1}{2} \sum_{j=1}^{+\infty} \frac{\free_j}{j} \Tr \left ( \ \frac{\rm i \sqrt{\etanew}}{N} \textbf{R}^T  \textbf{R} \underline L \right ) ^j   = \frac{1}{2} \sum_{j=1}^{+\infty} \frac{\free_j}{j} \Tr\left ( \ \frac{\rm i \sqrt{\etanew}}{N} \sum_{i=1}^N \vec{r}_i  \vec{r}_i^{\; T} \underline L  \right ) ^j  \nonumber \\
    & \quad = \frac{1}{2} \sum_{j=1}^{+\infty} \frac{\free_j}{j} \Tr \left ( {\rm i} \sqrt{\etanew} \int \mathrm{d} \vec y \, \mu (\vec y) \, \vec  y \vec y ^{\; T} \underline L \right ) ^j  = \frac{1}{2} \Tr \mathcal{H}_{\textbf{X}} \left ( {\rm i} \sqrt{\etanew} \int \mathrm{d} \vec{y} \, \mu (\vec{y}) \, \vec y\vec y ^{\; T} \underline L   \right ). \n
\end{align} 
With similar manipulations applied to the other factors in Eq.~\eqref{eq:step_2}, 
we derive
\begin{equation}
    \label{eq:step_3}
    Q_{[\alpha, \beta]}(n_{\pm}) \approx \int \mathcal{D} \mu \mathcal{D} \tilde \mu \, \ee^{N\mathcal{S}_n [\mu, \tilde \mu ; \underline \Lambda]},
\end{equation}
 where the action is given by
\begin{align}
    \label{eq:action}
    \mathcal{S}_n [\mu, \tilde \mu ; \underline \Lambda] = & -{\rm i} \int \mathrm{d} \vec{y} \, \mu (\vec{y}) \, \tilde \mu (\vec{y}) + \ln \int \mathrm{d} \vec{y} \,\ee^{-\frac{{\rm i}}{2} \vec{y}^{\, T} \underline \Lambda \vec{y} + {\rm i} \tilde \mu ( \vec{y})} \: \psi_a \left ( - \frac{1}{2} \vec{y}^{\; T} \underline L  \vec{y} \right ) \nonumber \\
    & + \frac{1}{2} \Tr \mathcal{H}_{\textbf{X}} \left ( {\rm i} \sqrt{\etanew} \int \mathrm{d} \vec{y} \, \mu (\vec{y}) \, \vec y \vec y ^{\, T}  \underline L\right )
    \; .
\end{align} 

As a check, we can inspect the limiting case in which $\newb$ belongs to the GOE, thus leading to the standard GRP model 
studied in Ref.~\cite{Venturelli_2023} within an analogous replica formalism.\footnote{The standard GRP model studied in Ref.~\cite{Venturelli_2023},
in which $\newb$ is a GOE matrix with average spectrum supported in $[-\sqrt{2},\sqrt{2}]$,
can be formally recovered here upon setting $\free_2=1/2$, see \cref{eq:H_series},  and identifying $\etanew=4\eta$.} 
In this case, the R-transform is simply $\mathcal{R}_{\rm GOE} (z) = z \Leftrightarrow \mathcal{H}_{\rm GOE} (z) = z^2/2$.  Plugging this into Eq.~\eqref{eq:action}, we indeed recover 
the action in Eq.~(73) of Ref.~\cite{Venturelli_2023}.

In the following, we will be
interested in the intermediate phase $1< \gamma<2$
of the phase diagram depicted in \cref{sec:model}. In this regime, the parameter $\etanew \propto N^{1-\gamma}$ is vanishingly small, which allows us to expand the second line of the action~\eqref{eq:action} up to order $\etanew$:
\begin{align}
    \label{eq:action_order_eta}
    &\mathcal{S}_n [\mu, \tilde \mu ; \underline \Lambda] =  -{\rm i} \int \mathrm{d} \vec{y} \, \mu (\vec{y}) \, \tilde \mu (\vec{y}) + \ln \int \mathrm{d} \vec{y} \, \ee^{-\frac{{\rm i}}{2} \vec{y}^{\, T} \underline \Lambda \vec{y} + {\rm i} \tilde \mu ( \vec{y})} \: \psi_a \left ( - \frac{1}{2} \vec{y}^{\; T} \underline L  \vec{y} \right ) \\
    & \qquad 
    + \frac{\rm i}{2}  \sqrt{\etanew} \free _1 \int \mathrm{d} \vec{y} \, \mu (\vec{y}) \,  \vec y ^{\, T}  \underline L \vec y - \frac{\etanew \free _2}{4}\int \mathrm{d} \vec{y} \,\mathrm{d} \vec{ w } \,  \mu (\vec{y}) \, \mu (\vec{w}) \, \left ( \vec y ^{\, T}  \underline L \vec w \right )^2 + o (\etanew )
    \; , \n
\end{align} 
\rev{where we used the expansion of $\mathcal{H}_{\textbf{X}}(z)$ in free cumulants given in \cref{eq:H_series}. We stress that the existence and finiteness of the free cumulants $\free_j$ is essential in order for this step to be justified; in particular, in the following we will assume $\free_2$ to be finite.}
For simplicity, we will \rev{further} assume that
$\free_1 = 0$ (see \cref{eq:H_series}),  which corresponds to requiring that the average spectral density of the matrix $\newb$ is centered, being
\begin{equation}
    \free_1 = \int \dd{\lambda} \lambda \,\rho_{\newb}(\lambda) = \frac{1}{N} \expval{\Tr \newb}.
\end{equation}
In this case, the action~\eqref{eq:action_order_eta} reduces, up to order $\etanew$, to
\begin{align}
    \mathcal{S}_n [\mu, \tilde \mu ; \underline \Lambda] = & -{\rm i} \int \mathrm{d} \vec{y} \, \mu (\vec{y}) \, \tilde \mu (\vec{y}) + \ln \int \mathrm{d} \vec{y} \, \ee^{-\frac{{\rm i}}{2} \vec{y}^{\, T} \underline \Lambda \vec{y} + {\rm i} \tilde \mu ( \vec{y})} \: \psi_a \left ( - \frac{1}{2} \vec{y}^{\; T} \underline L  \vec{y} \right ) 
    \nonumber\\
    &  - \frac{\etanew \free _2}{4}\int \mathrm{d} \vec{y} \,\mathrm{d} \vec{ w } \,  \mu (\vec{y}) \, \mu (\vec{w}) \, \left ( \vec y ^{\, T}  \underline L \vec w \right )^2 + o (\etanew )
    \; , 
        \label{eq:action_order_eta_2}
\end{align} 
which is the same as in the standard GRP model --- cfr.~Eq.~(73) of Ref.~\cite{Venturelli_2023}. This result implies that, on the scale of $\etanew$,  only the second free cumulant of $\newb$ matters in determining the full counting statistics.

The action~\eqref{eq:action_order_eta_2} will be the starting point of our calculation of the full counting statistics in \cref{sec:scaling-function}. Before doing so, however, we derive in the next Section an analogous action for the case of Wigner matrices.

\subsection{Wigner matrices}
\label{sec:wigner}

We now consider the case in which $\newb$ is a Wigner matrix, meaning that its entries~\eqref{eq:def_entries} are independent random variables drawn from the distribution $p_\zeta(x)$, not necessarily Gaussian. 
For convenience, we introduce 
the cumulant generating function\footnote{The case of
a standard GOE matrix with average spectrum supported in $[-\sqrt{2},\sqrt{2}]$ 
is obtained by choosing
$p_\zeta(x)$ to be Gaussian with zero mean ($\cum_1=0$) and variance $\cum_2= 1/2$. The corresponding cumulant generating function reads $F(t) = -\cum_2 t^2/2 = -t^2/4$. This is the case studied in Ref.~\cite{Venturelli_2023} within the standard GRP model (upon identifying $\etanew=4\eta $ therein).}
\begin{equation}
    F(t) \equiv \ln \int_{\mathbb R} \dd{x} \, p_\zeta(x) \mathrm \ee^{\mathrm i t x}
    = \sum_{j=1}^\infty \frac{\cum_j}{j!}(\mathrm i t)^j,
    \label{eq:def_cum}
\end{equation}
where $\cum_j$ are the (standard) cumulants of $p_\zeta(x)$.
Below we will only require the variance $\cum_2$ 
to be well defined.
This excludes, in particular, the case of Lévy matrices~\cite{biroli2021levy,safonova2025} --- we will further comment on this point in \cref{sec:discussion}.

To compute $Q_{[\alpha, \beta]}(n_{\pm})$ as defined in Eq.~\eqref{eq:Q_n_pm},
we follow the same steps as in Section~\ref{sec:free} that led to Eq.~\eqref{eq:step_1}. Here the crucial remark is that, thanks to the independence of the entries of $\newb$, one can immediately evaluate
\begin{align}
    \Biggl \langle \exp \left( \frac{{\rm i}\sqrt{\etanew}}{2} \sum_{i,j = 1}^N  m_{ij}  \Vec{r}_i^{\; T} \underline{L} \Vec{r}_j \right) \Biggr \rangle_{\newb} &= 
    \Biggl \langle \exp \left( \frac{{\rm i} \sqrt{\etanew}}{2} \sum_{i,j = 1}^N  m_{ij} \sum_{\rho=1}^n  r_{i\rho} \ell\ss r_{j\rho} \right) \Biggr \rangle_{\newb} \n\\
    &= \exp[\sum\ij F\left(\sqrt{\frac{\etanew}{2N}} \sum_{\rho=1}^n  r_{i\rho} \ell\ss r_{j\rho}\right) ]
    \, ,
\end{align}
in terms of the generating function $F(t)$ introduced in \cref{eq:def_cum}.
To derive its functional integral representation, we insert a decomposition of the identity in the form
\begin{equation}
    1 = \int \prod_{\rho=1}^n \dd{y_\rho} \delta (y_\rho -r_{i\rho})
    \; .
\end{equation}
In this way, we obtain
\begin{align}
    &\sum\ij F\left(\sqrt{\frac{\etanew}{2N}} \sum_{\rho=1}^n  r_{i\rho} \ell\ss r_{j\rho}\right) \n\\
    &
    \qquad 
    = \sum\ij \prod_{\rho=1}^n \int \dd{y_\rho} \delta (y_\rho -r_{i\rho}) \int \dd{w_\rho} \delta (w_\rho -r_{j\rho}) F\left( \sqrt{\frac{\etanew}{2N}} \sum_\rho y_\rho \ell\ss w_\rho  \right) \n\\
    &
    \qquad
    = N^2  \int \dd{\vec{y}} \dd{\vec{w}} \mu(\vec{y}) \mu(\vec{w}) 
    F\left( \sqrt{\frac{\etanew}{2N}}\vec{y}^{\; T} \underline L \vec{w}\right)
    \, ,
\end{align}
where we recognized the density $\mu(\vec{y})$ introduced in Eq.~\eqref{eq:mu-density}, and where $\vec y \in \mathbb R^n$ has components $y_\rho\in \mathbb{R}$.
Upon introducing
\begin{equation}
    \mathcal{K}(\vec{y}, \vec{w}) \equiv N \,F\left( \sqrt{\frac{\etanew}{2N}}\vec{y}^{\; T} \underline L \vec{w}\right) 
    \, ,
    \label{eq:M-def}
\end{equation}
and
inserting into Eq.~\eqref{eq:step_1}
the functional integral representation of the identity~\eqref{eq:identity}, we find that $Q_{[\alpha, \beta]}(n_{\pm})$
can again be expressed as a path integral over the fields $\mu$ and $\tilde \mu$ as in \cref{eq:step_3}, but with the action 
\begin{align}
    \label{eq:action-fix}
    \mathcal{S}_n [\mu, \tilde \mu ; \underline \Lambda] = & -{\rm i} \int \mathrm{d} \vec{y} \, \mu (\vec{y}) \, \tilde \mu (\vec{y}) 
    + \ln \int \mathrm{d} \vec{y} \,\mathrm \ee^{-\frac{{\rm i}}{2} \vec{y}^{\, T} \underline \Lambda \vec{y} + {\rm i} \tilde \mu ( \vec{y})} \: \psi_a \left ( - \frac{1}{2} \vec{y}^{\; T} \underline L  \vec{y} \right ) 
    \nonumber \\& 
    + \int \mathrm{d}\vec{y} \, \mathrm{d}\vec{w} \; \mu(\vec{y}) \mathcal K(\vec{y}, \vec{w})  \mu(\vec{w})
    \, .
\end{align} 
This action is quadratic in the field $\mu$ (whereas the one in \cref{eq:action} was not).
Upon setting
$\tilde \mu (\vec{y}) = \mathrm i \int \mathrm{d}\vec{w} \; \mathcal K(\vec{y}, \vec{w}) \varphi (\vec{w})$ and integrating over the field $\mu$, the action~\eqref{eq:action-fix} reduces to
\begin{align}
    \mathcal{S}_{n}[\varphi ;\underline \Lambda] =& - \frac14 \int \dd{\vec y} \dd{\vec w} 
    \, \varphi(\vec y) \mathcal K(\vec y,\vec w) \varphi(\vec w)  
    \nonumber\\
    &+ \ln \int \dd{\vec y} \, 
    \exp[-\frac{{\rm i}}{2}\vec y^{\; T} \underline \Lambda \vec y 
    -\int \dd{\vec w} \mathcal K(\vec y,\vec w) \varphi(\vec w) ] \, \psi_a\left(-\frac{1}{2} \vec y^{\; T}  \underline L  \vec y\right) .
    \label{eq:action_comp_varphi}
\end{align}

Again, we focus on the intermediate phase, $1< \gamma < 2$, and we expand the action for large $N$. This corresponds to expanding the function $\mathcal K(\vec{y}, \vec{w})$ in \cref{eq:M-def} as 
\begin{equation}
    \mathcal K(\vec{y}, \vec{w}) = 
    \mathrm i \cum_1 \sqrt{\frac{\etanew N}{2}} (\vec{y}^{\, T} \underline L \vec{w}) -
    \frac{\etanew\cum_2}{4} (\vec{y}^{\, T} \underline L \vec{w})^2 + o (\etanew).
    \label{eq:def_M_wigner}
\end{equation}
For a symmetric distribution, 
we have $\cum_1=0$ and thus
\begin{equation}
    \mathcal K(\vec{y}, \vec{w}) =  -
    \frac{\etanew\cum_2}{4} (\vec{y}^{\; T} \underline L \vec{w})^2 + o (\etanew) 
    \, .
    \label{eq:def_M_wigner_lead}
\end{equation}
As a result, at leading order in $\etanew$,
the action~\eqref{eq:action-fix} coincides with the one in \cref{eq:action_order_eta_2} for the mutually free case\footnote{We remark that the term $\propto \cum_1$ in the action~\eqref{eq:action-fix} for the Wigner case (with $\mathcal K(\vec y,\vec w)$ given in Eq.~\eqref{eq:def_M_wigner}) does not coincide with the term $\propto \free_1$
in the action~\eqref{eq:action_order_eta} for the mutually free case. This is not  worrying {\it per se}, since free cumulants are not the same as standard cumulants. Interestingly, the case of a GOE matrix 
falls into both categories, and corresponds to the choice $\free_2=\cum_2$, and $\free_n=\cum_n=0,\, \forall n\neq 2$ --- indeed, in this case the two actions in \cref{eq:action-fix,eq:action_order_eta} actually coincide.} --- which in turn coincides with the action of the standard GRP model (see Eq.~(D.5) in Ref.~\cite{Venturelli_2023}).

In the next Section, we will thus be able to compute
the CGF~\eqref{eq:CGF-def} simultaneously 
for both classes of matrices $\newb$ defined in~\cref{sec:model} (i.e.~the result will be exactly the same, at leading order, upon replacing $\cum_2 \leftrightarrow \free_2$).

\section{Full counting statistics}
\label{sec:scaling-function}

We now set out to compute the CGF~\eqref{eq:CGF-def}, which fully characterizes the number $I_N[\alpha,\beta]$ of eigenvalues in a given interval $[\alpha,\beta]$. In the following, we will choose a symmetric interval $[-E,E]$, and we will focus first on the case in which $E$ is of the order of the Thouless energy $E_T$ in \cref{eq:Thouless_energy}.
Our main result will be to express the CGF in the form
\begin{equation}
        \mathcal F_{[-E,E]} (s) = p_a(0) E_T\,  G\left(s, \frac{E}{E_T}\right) + \order{\etanew^2},
        \label{eq:scaling-function-def}
\end{equation}
where 
\begin{equation}
    E_T= \pi p_a(0) \cum_2 \etanew
    = \pi p_a(0) \cum_2 \nu^2 N^{1-\gamma} 
    \label{eq:thouless}
\end{equation}
is the Thouless energy, $\cum_2$ is the variance of the distribution of the off-diagonal entries of the matrix $\newb$, see \cref{eq:def_cum},  and $G$ is a universal scaling function which we determine below (see c.f.~\cref{eq_scaling-function-G}).
From their definition in \cref{eq:cumulants-general},
the scaled cumulants $\kappa_j\sim\mathcal O(N)$ of $I_N[-E,E]$ can then be identified by expanding $G(s,y)$ in powers of~$s$,
    \begin{equation}
        N \mathcal F_{[-E,E]} (s) =N p_a(0) E_T\,  G\left(s, \frac{E}{E_T}\right) + \order{\etanew^2}= -\kappa_1 s + \frac{s^2}{2} \kappa_2 +\order{s^3,\etanew^2}.
        \label{eq:cumulants}
    \end{equation}
The ratio of the first two cumulants defines the level compressibility~\eqref{eq:levelcomp-first}, which we characterize in \cref{sec:compressibility}. With this in hand,  we will then be able to obtain the large-deviation function $\Phi(k)$, defined as the Legendre transform of the CGF:
\begin{align}
        \label{eq:def_ldf}
        &\mathrm{Prob}\{I_N[-E,E]=kN p_a(0) E_T\} \simeq \mathrm \ee^{-N p_a(0) E_T \Phi(k)},\\
        &\Phi(k) = - \inf_{s}\{ks+G(s,E/E_T)\}. \label{eq:legendre:psi}
\end{align}

Finally, we show in \ref{app:E-O-1}
how the same method can be extended to obtain the CGF for energies $E\sim \mathcal{O}(1)$, within the same level of accuracy for large $N$. The result is reported in \cref{eq:CGF-E-O-1,eq:r-theta} below.

\subsection{Replica calculation of the scaled CGF}

In this Section, we present the calculation for the case in which $\textbf{M}$ is a Wigner matrix.
The corresponding result for orthogonally invariant matrices is identical
(as discussed in \cref{sec:Path-integral}) and follows immediately upon replacing $\cum_2 \to \free_2$.
We begin by considering only the leading-order term for small $\etanew$ in Eq.~\eqref{eq:def_M_wigner}, 
such that $\mathcal K(\vec{y}, \vec{w})$ in \cref{eq:M-def} reduces to the form given in Eq.~\eqref{eq:def_M_wigner_lead}.
We then introduce the symmetric matrix $\underline Q $ defined as 
\begin{equation}
    \int \dd{\vec w} (\vec{y}^{\, T} \underline L \vec{w})^2 \varphi(\vec w) = \vec{y}^{\, T} \underline Q \vec{y}
    \; ,
\end{equation}
whose matrix elements
read 
\begin{equation}
    q_{ik} = \sum_{j,l=1}^n \int \dd{\vec w}  \varphi(\vec w) \ell_{ij}\ell_{kl}w_j w_l = \ell_{ii} \ell_{kk} \int \dd{\vec w}  \varphi(\vec w) w_i w_k
    \; .
\end{equation}
In the second step we used that the matrix $\underline L$ 
defined in Eq.~\eqref{eq:def_lambda_L} is diagonal. 
(Unless otherwise stated, all sums in this Section run from $1$ to $n$.)
Using that $\ell_{ii}^2=1$, one can actually notice that
\begin{align}
    \int \dd{\vec y}\dd{\vec w} \varphi(\vec y)(\vec{y}^{\, T} \underline L \vec{w})^2 \varphi(\vec w) 
    &= \int \dd{\vec y}\varphi(\vec y)(\vec{y}^{\, T} \underline Q \vec{y}) = \sum_{ik}Q_{ik} \int \dd{\vec y}\varphi(\vec y) y_i y_k 
    \n\\
    &= \sum_{ik}Q_{ik}^2 \ell_{ii}\ell_{kk} = \Tr (\underline Q \underline L)^2
    \; .
\end{align}
This way we can rewrite the action~\eqref{eq:action_comp_varphi} in the more compact form 
\begin{align}
    &\mathcal{S}_{n}[\underline Q ;\underline \Lambda] = 
    \frac{\etanew \cum_2}{16} \Tr (\underline Q \underline L)^2+ \ln \mathcal Z_\varphi \; , \label{eq:action-Q}\\
    &\mathcal Z_\varphi= \int \dd{a} p_a(a) \int \dd{\vec y} \,\exp[-\frac{1}{2}\vec y^{\; T} \left( \mathrm i \underline \Lambda  -
    \frac12
    \etanew \cum_2 \underline Q - \mathrm i a \underline L \right)\,\vec y]  \; . \label{eq:Zphi}
\end{align}
Our aim is to optimize the action with respect to the matrix elements $q_{ij}$, and then obtain the scaled CGF from Eqs.~\eqref{eq:CGF_replica}~and~\eqref{eq:step_3} as
\begin{equation}
    \mathcal F_{[-E,E]} (s) \asymp \lim_{\varepsilon\to 0^+}  \eval{\mathcal{S}_{n}[\underline Q ;\underline \Lambda]}_{n_\pm = \pm  \frac{\mathrm i s}{\pi}} 
    \; .
    \label{eq:F-S}
\end{equation}
Guided by the replica-symmetric solution found in Ref.~\cite{Venturelli_2023}, we now make the assumption that the matrix $\underline Q$ 
shares the same block-diagonal form as the matrices $\underline \Lambda$ and $\underline L$
in \cref{eq:def_lambda_L},
namely $\underline{Q} = \mathrm{diag}(
q_1 1_{n_+},\,
  q_2 1_{n_+} ,\,
 q_3 1_{n_-} ,\, q_4  1_{n_-} )$. 
 The corresponding four saddle-point equations, found by imposing 
 \begin{equation}
     \pdv{\mathcal{S}_{n}[\underline Q ;\underline \Lambda]}{q_k} = 0
     \; ,
 \end{equation}
imply that 
\begin{equation}
    q_k = - \frac{2}{\mathcal Z_\varphi} \int \dd{a} p_a(a) \int \dd{\vec y} \, y_k^2 \, \exp[-\frac{1}{2}\vec y^{\; T} \left( \mathrm i \underline \Lambda  - 
    \frac12
    \etanew \cum_2 \underline Q - \mathrm i a \underline L \right) \vec y]
    \; .
    \label{eq:spe-q}
\end{equation}
To find the scaling function~\eqref{eq:scaling-function-def}, it turns out that it is actually sufficient to solve the saddle-point equation~\eqref{eq:spe-q} at leading order for small $\etanew$. To this end, we first specialize the matrix $\underline \Lambda$ in Eq.~\eqref{eq:def_lambda_L} to the symmetric interval $[\alpha,\beta]=[-E,E]\equiv [-x\etanew,x\etanew]$, with $x\sim \order{1}$, and we decompose it as 
\begin{equation}
    \underline \Lambda \equiv x\etanew\, \underline{J} - \mathrm i \varepsilon\, 1_{n}
    \; ,
    \label{eq:decomp-Lambda}
\end{equation}
upon introducing the matrix
\begin{equation}
\label{eq:def_lambda_J}
    \underline{J}  = \mqty(\dmat{
 -1_{2n_+}\!\!\!\!,  1_{2n_-} })
 \; ,
\end{equation} 
where we recall that $n=2(n_++n_-)$. 
We then replace $q_k=q_k\z+\order{\etanew}$ into the saddle-point equation~\eqref{eq:spe-q} and discard all terms of $\order{\etanew}$, finding at lowest order 
\begin{equation}
    q_k\z = - \frac{2}{\mathcal Z_\varphi\z} \int \dd{a} p_a(a) \int \dd{\vec y} \, y_k^2 \, \exp[-\frac{1}{2}\vec y^{\; T} \left( \varepsilon\, 1_{n} - \mathrm i a \underline L \right)\,\vec y] 
    \; ,
    \label{eq:spe-q-0}
\end{equation}
where
\begin{align}
    \label{eq:Z0}
    \mathcal Z_\varphi\z&= \int \dd{a} p_a(a) \int \dd{\vec y} \exp[-\frac{1}{2}\vec y^{\; T} \left( \varepsilon\, 1_{n} - \mathrm i a \underline L \right)\,\vec y] \n\\
    &=\int \dd{a} p_a(a)  (2\pi)^\frac{n}{2} (\varepsilon^2+a^2)^{-\frac{n}{2}} \xrightarrow[n_\pm \to \pm \mathrm i s/\pi]{} 1
    \; . 
\end{align}
Similarly,
\begin{align}
    &\int \dd{a} p_a(a) \int \dd{\vec y} y_k^2 \exp[-\frac{1}{2}\vec y^{\; T} \left( \varepsilon\, 1_{n} - \mathrm i a \underline L \right)\,\vec y] 
    =\int \dd{a} p_a(a)  (2\pi)^\frac{n}{2} \frac{(\varepsilon^2+a^2)^{-\frac{n}{2}}}{\varepsilon-\mathrm i a \ell_{kk}}  \n\\
    &\xrightarrow[n_\pm \to \pm \mathrm i s/\pi]{} \int \dd{a} \frac{p_a(a)}{\varepsilon-\mathrm i a \ell_{kk}}
    \; .
\end{align}
Using that $\ell_{kk}=\pm 1$, we can now rewrite
\begin{align}
    & \int \dd{a} \frac{p_a(a)}{\varepsilon-\mathrm i a \ell_{kk}} = \mathrm i \ell_{kk} \int \dd{a} \frac{p_a(a)}{a+\mathrm i \varepsilon \ell_{kk}} 
    \nonumber\\
    & 
    \qquad\qquad\quad\;\; \xrightarrow[\varepsilon \to 0^+]{} \mathrm i \ell_{kk} \left[ -\mathrm i \pi \ell_{kk} p_a(0) + \mathrm{P.V.} \int \dd{a} \frac{p_a(a)}{a} \right],
\end{align}
where we used the Sokhotski--Plemelj formula~\cite{Livan_2018}. 
For a symmetric distribution $p_a(a)$, the second term in square brackets vanishes, and from Eqs.~\eqref{eq:spe-q-0}~and~\eqref{eq:Z0} we find the very simple result
\begin{equation}
    q_k\z = - 2\pi p_a(0) \ell_{kk}^2 = - 2\pi p_a(0) \equiv - q
    \; ,
\end{equation}
with $q>0$. Then,
\begin{equation}
    \Tr (\underline Q \underline L)^2  = \sum_{ik}Q_{ik}^2 \ell_{ii}\ell_{kk} = \sum_{i=1}^n q^2 
    \xrightarrow[n_\pm \to \pm \mathrm i s/\pi]{} 0
    \; ,
\end{equation}
where we used the optimized solution $\underline Q = -q 1_{n} + \order{\etanew}$.
The action~\eqref{eq:action-Q} now reduces to
\begin{equation}
    \mathcal{S}_{n}[\underline Q ;\underline \Lambda] = \ln \mathcal Z_\varphi\o +\mathcal{O}(\etanew^2)
    \; ,
    \label{eq:S-lnZ}
\end{equation}
up to leading order, 
where $\mathcal Z_\varphi\o $ is obtained by inserting the leading-order solution for $\underline Q$ into 
$\mathcal Z_\varphi$ given by Eq.~\eqref{eq:Zphi} (but without discarding the other terms of $\order{\etanew}$, as we did instead
to obtain $\mathcal Z_\varphi\z$ in \cref{eq:Z0}), so that $\mathcal Z_\varphi= \mathcal Z_\varphi\o[1+\mathcal{O}(\etanew^2)]$. 
It proves convenient to use again the decomposition~\eqref{eq:decomp-Lambda} of the matrix $\underline \Lambda$, and to change variables as $\vec z= \vec y \sqrt{\etanew} $ and $b=a/\etanew$ to obtain
\begin{align}
    \mathcal Z_\varphi\o=  \etanew^{1+n}\int \dd{b} \, p_a(\etanew b) \int \dd{\vec z} \, \exp[-\frac{1}{2}\vec z^{\; T} \left( \frac{\varepsilon}{\etanew}\, 1_{n} +\mathrm i x \underline{J} - 
    \frac12
    \cum_2 \underline Q - \mathrm i b \underline L \right) \vec z \, ]  
    \; .
\end{align}
Upon inserting the leading-order solution $\underline Q = -q 1_{n} + \order{\etanew}$ and computing the Gaussian integral over $\vec z$, in the limit $\varepsilon/\etanew \to 0^+$ we find\footnote{We note that the regulator $\varepsilon$ must actually tend to zero faster than $\etanew$, when $N$ becomes large. This is reminiscent of the discussion on symmetry breaking found in Ref.~\cite{Cavagna_2000}.} 
\begin{align}
    \mathcal Z_\varphi\o &=  \etanew^{1+n}\int \dd{b} \, p_a(\etanew b) (4\pi)^\frac{n}{2} \left\lbrace
    \left[ \cum_2 q -2\mathrm i (x+b)  \right]
    \left[  \cum_2 q -2\mathrm i (x-b)  \right]
    \right\rbrace^{-\frac{n_+}{2}} \n\\
    &\qquad\qquad \qquad\qquad \quad \;\;\, \times
    \left\lbrace
    \left[ \cum_2 q +2\mathrm i (x+b)  \right]
    \left[\cum_2 q +2\mathrm i (x-b)  \right]
    \right\rbrace^{-\frac{n_-}{2}}
    \n\\
    &\!\!\!\! 
    \xrightarrow[n_\pm \to \pm \mathrm i s/\pi]{} \etanew \int \dd{b} p_a(\etanew b) \exp{\frac{\mathrm i s}{2\pi} \ln\frac{\left[ \cum_2 q+2\mathrm i (x+b)  \right]
    \left[\cum_2 q+2\mathrm i (x-b)  \right]}{\left[ \cum_2 q-2\mathrm i (x+b)  \right]
    \left[  \cum_2 q-2\mathrm i (x-b)  \right]}}
    \; .
    \label{eq:Zphi-step}
\end{align}
For consistency, we will now only retain the leading-order contribution in $\etanew$ by expanding $p_a(\etanew b)= p_a(0)+\order{\etanew}$, which is actually the key to obtaining a scaling form that is independent of the particular choice of $p_a(a)$ \rev{(up to a rescaling by $p_a(0)$)}. This must however be handled with care, because the remaining integral diverges for $s=0$ if one naively inserts the expansion of $p_a(\etanew b)$ above. To regularize it, we note that
\begin{equation}
    1 = \int \dd{a} \, p_a(a) =
    \etanew \int \dd{b} \, p_a(\etanew b)
    \; ,
\end{equation}
and hence
\begin{align}
    \etanew \int \dd{b} \, p_a(\etanew b) \ee^{\mathrm i s (\dots)} &= 1 + \etanew \int \dd{b} \, p_a(\etanew b) \left[\ee^{\mathrm i s (\dots)}-1\right] \n\\
    &= 1 + \etanew p_a(0) \int_{\mathbb{R}} \dd{b} \,  \left[\ee^{\mathrm i s (\dots)}-1\right] +\order{\etanew^2}
    \; ,
\end{align}
where the last integral is now convergent also for $s=0$.
This way we get from \cref{eq:Zphi-step}, upon expanding $\ln(1+\etanew z) = \etanew z + \order{\etanew^2}$,
\begin{equation}
    \lim_{n_{\pm} \to \pm is/\pi} \ln \mathcal Z_\varphi\o = \etanew p_a(0) \int_{\mathbb{R}} \dd{b}  \left[\ee^{\mathrm i s (\dots)}-1\right] +\order{\etanew^2}
    \; .
    \label{eq:lnZ-approx}
\end{equation}
Next, we write in polar form $(\cum_2 q/2 +\mathrm i x )= r \mathrm \ee^{\mathrm i \theta}$, where
\begin{equation}
    r^2 = (1+y^2) \left(\frac{\cum_2 q}{2}\right)^2, \qquad \theta = \arctan(y)\in [-\pi/2,\pi/2]
    \; , 
\end{equation}
with
\begin{equation}
    y = \frac{x}{\cum_2 q/2} =  \frac{x}{\cum_2 \pi p_a(0)} = \frac{E}{E_T}>0
    \; ,
\end{equation}
and where we recognized the Thouless energy~\eqref{eq:thouless}.
Using the identity $2\mathrm i \arctan(z) = \ln[(1-\mathrm i z)/(1+\mathrm i z)]$ and changing variables as
$v=b/r$, we can rewrite
\begin{align}
    & \lim_{n_{\pm} \to \pm is/\pi} \ln \mathcal Z_\varphi\o = p_a(0) E_T\sqrt{1+y^2}
    \int_{-\infty}^\infty \dd{v} \left\lbrace \exp[ -\frac{s}{\pi} \arctan(\frac{\sin(2\theta)}{v^2+\cos(2\theta)})]-1\right\rbrace 
    \quad
    \nonumber\\
    & 
    \qquad\qquad\qquad\quad\, +\order{\etanew^2} 
    \; .
    \end{align}
Since for $y>0$ the angle $\theta$ is in 
the interval 
$\theta \in [0,\pi/2]$, this expression can be further simplified by rewriting $\sin(2\theta) = 2y/(1+y^2)$ and $\cos(2\theta) = (1-y^2)/(1+y^2)$. 
Upon changing variables as $u=v \sqrt{1+y^2}$,
we can then use \cref{eq:S-lnZ,eq:lnZ-approx,eq:F-S} to obtain
\begin{align}
    \mathcal F_{[-E,E]} (s) = \eval{\mathcal{S}_{n}[\underline Q ;\underline \Lambda]}_{n_\pm = \pm  \frac{\mathrm i s}{\pi}}  =p_a(0) E_T\, G(s,E/E_T)+ \order{\etanew^2} 
    \; ,
    \label{eq:scaled-CGF}
\end{align}
where $y=E/E_T >0$,
with $E_T$ the Thouless energy~\eqref{eq:thouless}, and
with the scaling function
\begin{align}
    G(s,y)
    = \int_{-\infty}^\infty \dd{u} \left\lbrace \exp[ -\frac{s}{\pi} \arctan(\frac{2y}{u^2+1-y^2})]-1\right\rbrace
    \; .
    \label{eq_scaling-function-G}
\end{align}
Equations~\eqref{eq:scaled-CGF}~and~\eqref{eq_scaling-function-G} represent the main result of this Section. 
For $E=E_T$, the \rev{convex} function $G(s,1)$ decreases monotonically from $+\infty$ to $-\infty$, and crosses zero when $s=0$.

In passing, we note that the first discarded term of $\order{\etanew^2}$ in \cref{eq:scaled-CGF} represents the leading correction to the result, larger (for $\gamma<2)$ than the 1-loop correction of $\order{\etanew/N}$ coming from the Gaussian fluctuations around the saddle point~\cite{Venturelli_2023}.

Finally, a completely analogous calculation leads to the full counting statistics for generic energies $E$ --- i.e.~not necessarily of $\order{\etanew}$, with the only requirement that $E$ is larger than the mean level spacing $\delta_N\propto N^{-1}$, otherwise the saddle-point construction breaks down~\cite{Venturelli_2023}. We defer the derivation to \ref{app:E-O-1}, while we report here only the final result:
\begin{align}
    \label{eq:CGF-E-O-1}
    \mathcal F_{[-E,E]} (s) =& -\frac{s\etanew \cum_2}{4\pi}\Im{\mathfrak q^2} +\ln \int_{-\infty}^\infty \dd{a} p_a(a) \exp[ -\frac{s}{\pi} \arctan \left( \frac{ \sin 2\theta}{a^2r^2 + \cos 2\theta} \right) ]\n\\
    &+\mathcal{O}(\etanew^2), 
\end{align}
where $\mathfrak q = 2\mathrm i \mathcal{G}_a(E+\mathrm i 0^+)$ in terms of the Cauchy--Stieltjes transform \rev{$\mathcal G_a(z)$} of the distribution $p_a(a)$ (see \cref{eq:resolvent}), while the parameters $r$ and $\theta$ are implicitly defined~by 
\begin{equation}
     (\etanew\cum_2 \mathfrak q/2 -\mathrm i E)^{-1} \equiv r\mathrm \ee^{\mathrm i \theta}.
     \label{eq:r-theta}
\end{equation}
We note that the CGF~\eqref{eq:CGF-E-O-1}
depends explicitly on $p_a(a)$, whereas the particular choice of $\textbf{M}$ \rev{in \cref{eq:RPdef}} only enters through a single parameter, namely the variance $\cum_2$ of $p_\zeta(x)$ (in the Wigner case, while we recall that $\cum_2$ has to be replaced by the second free cumulant $\free_2$ in the orthogonally invariant case).

\subsection{Level compressibility and large-deviation function}
\label{sec:compressibility}

Upon expanding $G(s,y)$ in powers of $s$ as in Eq.~\eqref{eq:cumulants}, we can use \cref{eq_scaling-function-G} to recover the first few moments of $I_N[-E,E]$. First, we note that $G(s=0,y)=0$ and $G(s,y=0)=0$, as it should be by construction (see \cref{eq:CGF-def}). Next,
\begin{equation}
    \frac{\kappa_1}{N p_a(0) E_T} = -\eval{\partial_s G(y,s)}_{s=0} = \frac{1}{\pi} \int_{-\infty}^\infty \dd{u} \, \arctan(\frac{2y}{u^2+1-y^2}) = 2y.
    \label{eq:k1}
\end{equation}
This can be proven by differentiating the integrand with respect to $y$, factorizing 
$(u^2+1-y^2)^2+4y^2=[(u+y)^2+1][(u-y)^2+1]$, computing the integral with respect to $u$, and finally integrating again with respect to $y$. The result 
\begin{equation}
\label{eq:mean:I_N}
    \kappa_1 = 2Np_a(0) E
\end{equation}
is very intuitive, since for $E=x \etanew \ll 1$ one has $\rho_N(\lambda) \simeq p_a(\lambda)$ and thus
\begin{equation}
    \label{eq:exp_I_N}
    \expval{I_N[-E,E] } = N \int_{-E}^E \dd{\lambda} \, \rho_N(\lambda)  \simeq 2Np_a(0) E.
\end{equation}
Next, we inspect
\begin{align}
    \frac{\kappa_2}{N p_a(0) E_T} &= \eval{\partial_s^2 G(y,s)}_{s=0} = \frac{1}{\pi^2} \int_{-\infty}^\infty \dd{u} \left[\arctan(\frac{2y}{u^2+1-y^2})\right]^2 \n\\
    &= \frac{2}{\pi} \left[2y \arctan(y)-\ln(1+y^2)  \right].
    \label{eq:k2}
\end{align}
(\rev{Evaluating this integral} 
is quite non-trivial~\cite{stack}, but it can be easily checked e.g.~using \texttt{Mathematica}.)
Taking the ratio of \cref{eq:k2,eq:k1} yields the level compressibility
\begin{equation}
    \chi_T(y) = \frac{\kappa_2(y)}{\kappa_1(y)} = 
    \frac{1}{\pi y} \left[2y \arctan(y)-\ln(1+y^2)  \right],
    \label{eq:chi_universal}
\end{equation}
i.e.~the scaling function first found in Ref.~\cite{Venturelli_2023} for the case of the standard GRP model (see also~\cite{Delapalme2026}).

Finally, we can obtain the large-deviation function $\Phi(k)$ defined in Eq.~\eqref{eq:def_ldf} by
computing numerically the Legendre transform of $G(s,y)$ with respect to $s$. The result, together with a comparison to large-deviation simulations, is shown in 
Figure~\ref{fig:counting:rare}(b).
One can also check that the following asymptotic behaviors hold, for $y=1$:
\begin{equation}
\label{eq:rate:limits}
\Phi(k) \sim
    \begin{cases}
        2/k, & \text{for } k\to 0 \; , \\
        2k \ln k, & \text{for } k\to \infty \; .
    \end{cases}
\end{equation}

\section{Numerical tests and discussion}
\label{sec:numerics}

In this Section we present numerical results, obtained from exact diagonalization of random matrices.

First, we show results for the full counting statistics, where in particular we address  large-deviation properties of the distribution for medium-sized matrices by using specific rare-event algorithms, which are outlined in the following subsection. 

Furthermore, to test the universality of the level statistics around the Thouless energy, we measure the level compressibility for distinct random matrices, belonging to either of the two classes delineated in \cref{sec:model}, and we compare it to the scaling formula~\eqref{eq:chi_universal}.
These results are obtained through exact diagonalization of large random matrix samples,
using the algorithm \texttt{eigvals} from the \texttt{LinearAlgebra} package in Julia. 

\subsection{Large-deviation properties of the counting statistics}
\label{sec:counting:rare}

To measure the number $I_N$ of eigenvalues in the interval $[-E,E]$ we can 
directly sample matrices $\textbf{H}$ according to the original
statistics as given by Eq.~(\ref{eq:RPdef}), with densities here denoted as $Q(\textbf{H})$, and 
diagonalize the matrices. Then we simply count, by considering all eigenvalues, the number $I_N=I_N(\textbf{H})$ of eigenvalues which are located inside the interval. 
By performing  this direct sampling for many matrices, say $10^6$, 
we can obtain histograms of $I_N$ approximating $P(I_N)$ within a range of probabilities down to about $10^{-6}$, which is rather limited. 

In order to address much smaller probabilities, we apply specific large-deviation algorithms~\cite{dembo2010,les_houches2024}. The main idea is not to sample the matrices
according to the original densities $Q(\textbf{H})$, but according to a \emph{biased} (or \emph {tilted}) statistics $Q_{\Theta}(\textbf{H})$,
which is designed to shift the sampling into the tails of the distribution. A rather common bias is the exponential bias $\sim e^{-I_N(\textbf{H})/\Theta}$~\cite{align2002}, with $\Theta$ being a \emph{pseudo temperature}, which controls the bias. We apply this bias here, i.e.~the probability density for a matrix is given by
\begin{equation}
\label{eq:prob:biased}
    Q_{\Theta}(\textbf{H}) = \frac{1}{\mathcal Z(\Theta)} Q(\textbf{H}) 
    e^{-I_N(\textbf{H})/\Theta}\,,
\end{equation}
with normalization $\mathcal Z(\Theta)=\int_{\textbf{H}} d\textbf{H} \,Q(\textbf{H}) 
    e^{-I_N(\textbf{H})/\Theta}$. One can show easily~\cite{align2002} that
    this leads to a statistics of the number of eigenvalues $I_N$ as given by
    \begin{equation}
    \label{eq:prob:theta:IN}
        P_{\Theta}(I_N)= \frac{1}{\mathcal Z(\Theta)} e^{-I_N/\Theta} P(I_N)\, ,
    \end{equation}
i.e.~modified probabilities with respect to the desired probabilities $P(I_N)$.

The main idea is to sample the biased distribution for different suitably chosen values of $\Theta$, which will generate values of $I_N$ in different ranges of the support. For example, high values of $|\Theta|$ will change the probabilities only slightly, while very small and positive values shift the sampling to small values of $I_N$. Larger than typical values of $I_N$ can be obtained from negative values of $\Theta$. 

To achieve  the actual sampling according to $Q_{\Theta}(\textbf{H})$ we use a Markov chain Monte Carlo (MCMC) approach~\cite{newman1999}, where the configurations in the Markov chain are matrices $\textbf{H}(t)$ for $t=0,1,\ldots$. As start configuration, we 
take a random matrix sampled from the original distribution $Q(\textbf{H})$. To generate the Markov chain, we use the Metropolis algorithm. This means, given the current configuration $\textbf{H}(t)$, which exhibits $I_N(t)=I_N(\textbf{H}(t))$ eigenvalues in the interval $[-E,E]$, we generate a \emph{trial} matrix $\textbf{H}'$ by copying $\textbf{H}(t)$ and redrawing a number $n_{\rm c}$ of entries according to the original statistics. This realizes the $Q(\textbf{H})$ factor in Eq.~(\ref{eq:prob:biased}). For the trial matrix the number of eigenvalues $I_N'=I_N(\textbf{H}')$ is obtained. At this stage, the trial configurations is \emph{accepted}, i.e.~$\textbf{H}(t+1)=\textbf{H}'$, with the Metropolis probability
\begin{equation}
    p_{\rm acc} = \min \left\{ 1, e^{-(I_N'-I_N(t))/\Theta)} \right\}\,.
\end{equation}
Otherwise, with probability $1-p_{\rm acc}$, the current configuration is kept, i.e.~$\textbf{H}(t+1)=\textbf{H}(t)$. This Metropolis part realizes
the $e^{-I_N(\Theta)}$ factor in Eq.~(\ref{eq:prob:biased}). Note that the number $n_{\rm c}$ of changed values, to ensure efficient sampling, will depend on $\Theta$. The larger $|\Theta|$ is, the more entries can be changed while still accepting a sufficiently large fraction, about 50\%, of the trial matrices.

After discarding the initial part of the MCMC simulations, i.e.~waiting for \emph{equilibration}, one can obtain histograms of the sampled values $I_N$, which approximate the distributions $P_{\Theta}(I_N)$. Then, using Eq.~(\ref{eq:prob:theta:IN}), one obtains the (approximate) desired probabilities from
\begin{equation}
\label{eq:rescale:prob}
P(I_N)= \mathcal Z(\Theta) e^{I_N/\Theta} P_{\Theta}(I_N)
\; .
\end{equation}
The normalization factors $\mathcal Z(\Theta)$ are not known a priory, but it should be noted that their ratios $\mathcal Z(\Theta_1)/ \mathcal Z(\Theta_2)$ can be determined (see Sec 4.3 in~\cite{les_houches2024} or~\cite{align2002}) by equating the right-hand sides of Eq.~(\ref{eq:rescale:prob}) for two different temperature values $\Theta_1$ and $\Theta_2$. 
With the data obtained from the simulations, this can be done approximately by considering close-by values of $\Theta_1, \Theta_2$ such that the two corresponding sets of sampled values of $I_N$ overlap, and requiring that the pairs of estimated probabilities for $P(I_N)$ obtained from the simulations of $\Theta_1$ and $\Theta_2$ agree as much as possible, i.e.~minimizing the mean-squared difference. 
This yields an estimate for $\mathcal Z(\Theta_1)/\mathcal Z(\Theta_2)$.
By performing this step-wise for neighboring pairs of $\Theta$, one basically ``glues'' the different contributions of the distribution together.
The overall normalization to one yields the final missing equation to determine all values of $\mathcal Z(\Theta)$. For details see the literature~\cite{align2002,les_houches2024}.
To achieve this glueing, one can also use the multi-histogram approach~\cite{Ferrenberg_1989}, which is implemented in a convenient tool~\cite{werner2022}.  

Such large-deviation approaches have been successfully applied  to a wide range of problems,
including the properties of random graphs~\cite{diameter2018,werner2024}, work distributions for Ising models~\cite{work_ising2014}, the calculation of partition functions~\cite{partition2005}, sequence alignment 
statistics~\cite {align2002}, traffic model properties~\cite{nagel_schreckenberg2019}, 
and the height distributions of the Kardar-Parisi-Zhang equation~\cite{kpz2018}, among others.
    
We have implemented the large-deviation code in C and used the 
GNU Scientific Library\cite{gsl2006} to diagonalize the matrices.
The number of pseudo temperatures ranges between 4 for $N=20$ and 12 for $N=2000$. While the runs for small matrices finished within minutes, we needed 80 cores being run for more than one week to obtain the data for $N=2000$.

\begin{figure}[t]
    \centering
    \includegraphics[width=0.42\linewidth]{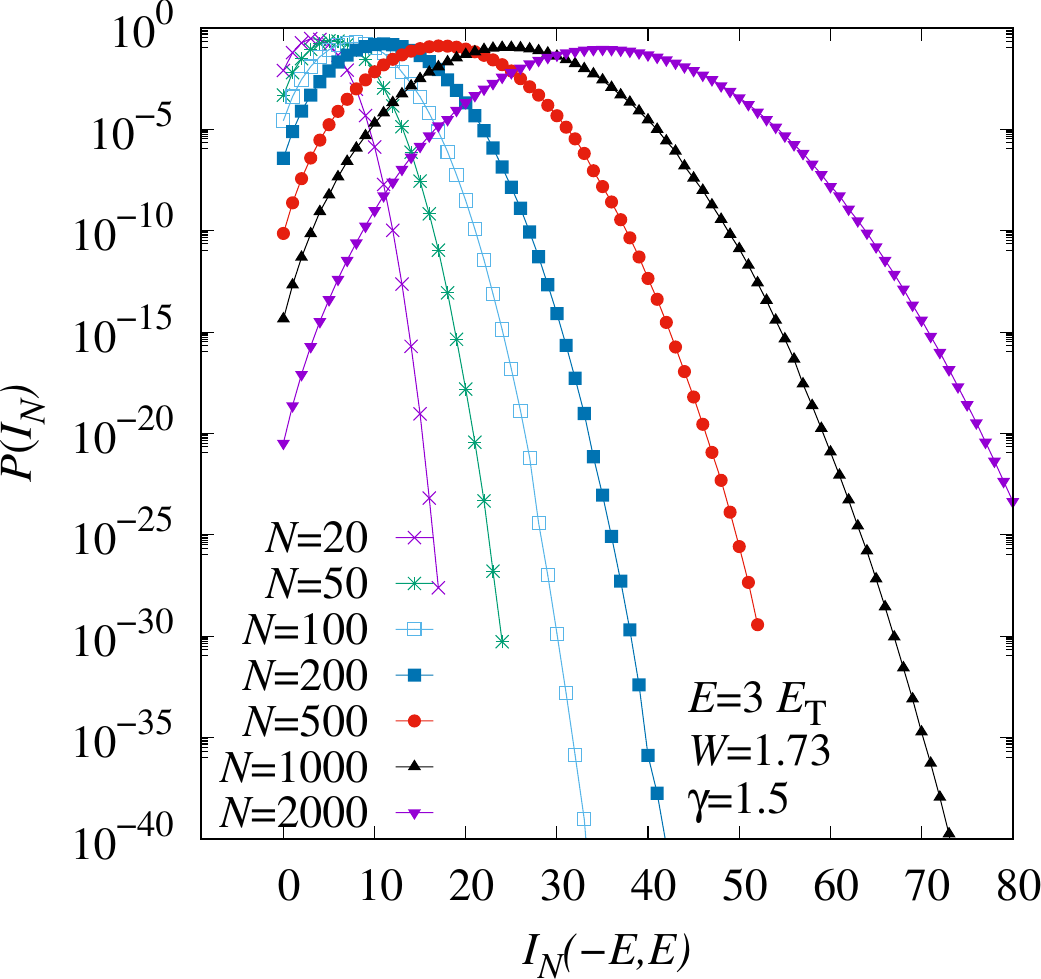}
    \hspace{20pt}
    \includegraphics[width=0.4\linewidth]{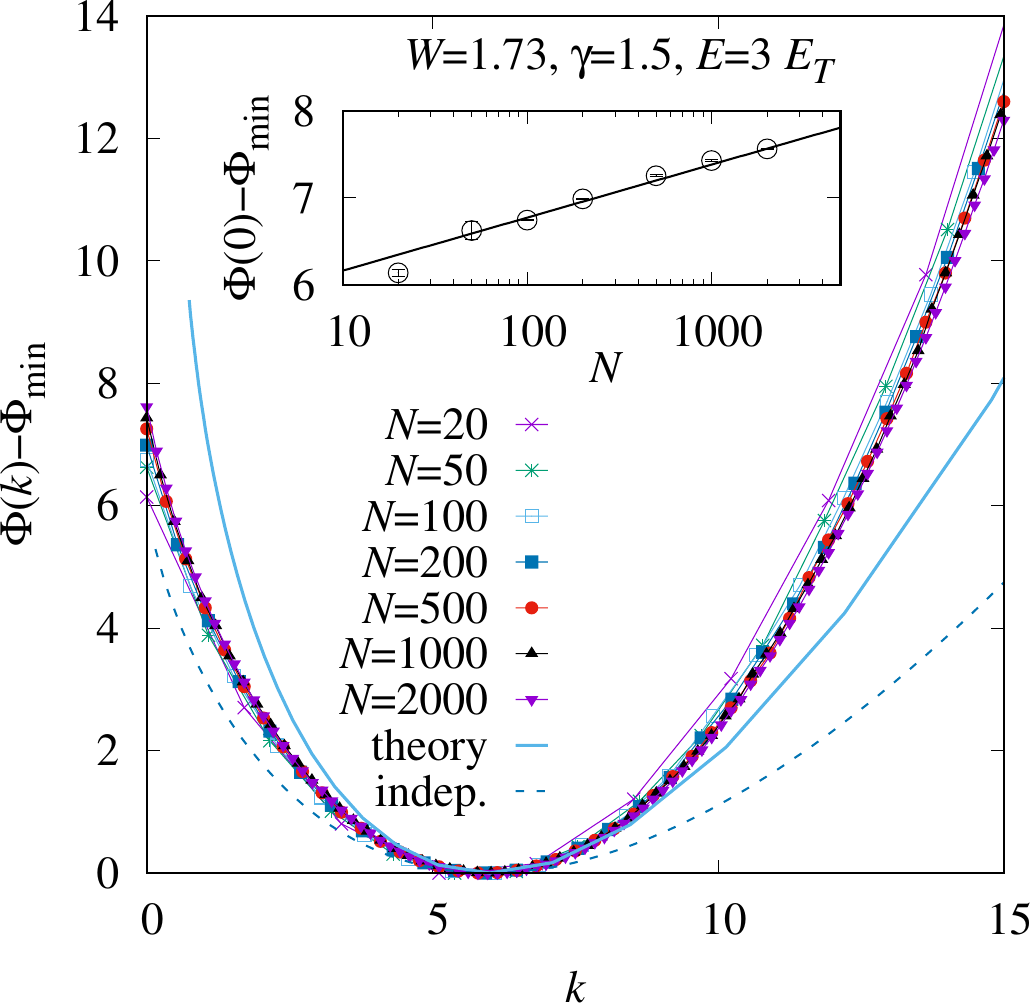}
    \put(-400,10){(a)}
    \put(-190,10){(b)}
    \caption{(a) The distribution of the full counting statistics 
    $I_N[-E,E]$ for $W=\sqrt{3}$, $\nu=1$, $\gamma=1.5$, $E=3 E_T$,
    and matrix sizes $N=20$, 50, 100, 200, 500, 1000 and 2000.
     (b) The corresponding shifted rate function $\Phi(k)-\Phi_{\min}$ as a function of the counting statistics $k=I_N/[Np_a(0)E_T]$, rescaled according to Eq.~(\ref{eq:def_ldf}). The solid line shows the theoretical prediction, while the dashed line the rate function obtained for independent levels from Eq.~(\ref{eq:psiinf}). The inset shows the behavior at $k=0$ as a function of the matrix size $N$, together with a fit to Eq.~(\ref{eq:log_fit}).
    }
    \label{fig:counting:rare}
\end{figure}

In Fig.~\ref{fig:counting:rare}(a) we show the distributions of the counting statistics $I_N(-E,E)$ obtained for $\nu=1$, $\gamma=1.5$, $p_a(a)$ uniform in $[-W,W]$ with $W=\sqrt{3}$, and $E=3E_T$ for different system sizes $N=20$, 50, 100, 200, 500, 1000 and 2000. 
The matrix $\textbf{M}$ in \cref{eq:RPdef} has been chosen to belong to the GOE as in~\cite{Venturelli_2023} --- equivalently, $\textbf{M}$ is a Wigner matrix with Gaussian entries of variance $\cum_2=1/2$.
Thanks to the application of the large-deviation  algorithm, probabilities as small as $10^{-40}$ can be reached. With growing size $N$, one observes a shift of the distributions to the right, as described by the first moment given by Eq.~(\ref{eq:mean:I_N}). The distributions also broaden slightly, 
as described by the second moment stated in Eq.~(\ref{eq:k2}).

The leading finite-size behavior of the first two moments is taken into account in Fig.~\ref{fig:counting:rare}(b), where we show the corresponding empirical 
rate function as a function of the scaled counting statistics $k$, corresponding to the definition in Eq.~(\ref{eq:def_ldf}). 
We have shifted the empirical rate functions by subtracting the minimum values $\Phi_{\min}$, corresponding to the maximum probability,  attained near the $N\to\infty$ minimum position, which is
$k_{\min}=\kappa_1/[Np_a(0)E_T]=6$ for $E=3E_T$, see \cref{eq:def_ldf,eq:mean:I_N}.
  Note that, in the limit $N\to\infty$, $\Phi(k_{\min})=0$ 
must hold because otherwise the probability distribution would not be normalized. 
Thus, this shift removes one finite-size correction. 
Near the minimum, a very good data collapse for different matrix sizes is visible, justifying the shift by subtracting the finite-size values of $\Phi_{\min}$. The collapsed data agrees here perfectly with the analytical result,
 which was obtained by numerical Legendre transformation (see \cref{eq:legendre:psi}) of \cref{eq_scaling-function-G}.
But for smaller values of $k\to 0$, a disagreement becomes visible --- in particular, Eq.~(\ref{eq:rate:limits}) predicts a divergence of $\Phi(0)$.
To analyze this growth, we have evaluated the shifted rate function at $k=0$ as a function of the matrix size $N$, as shown in the inset of  Fig.~\ref{fig:counting:rare}(b).
Indeed, $\Phi(0)-\Phi_{\min}$ increases logarithmically with $N$. Note that this growth is not a mere artefact of subtracting $\Phi_{\min}$, because $\Phi(0)$ actually decreases for small values of $N$ but then starts to increase again.
To analyze this growth quantitatively, we have fitted for $N\ge 50$ a logarithmic function
\begin{equation}
f(N)=f_0+f_1 \ln(N)
\label{eq:log_fit}
\end{equation}
for $N\ge 50$ to the data (yielding $f_0=5.6(7)$ and $f_1=0.26(1)$), with a fair agreement. Thus, it is possible that $\Phi(0)$ actually diverges, 
as predicted by the theory. Still, the divergence as a function of $N$ is extremely slow, such that we cannot rule out a different behavior  with no divergence of $\Phi(0)$. As usual with logarithmic divergences, we can also fit a power-law convergence to a constant to the data.  This slow finite-size behavior also means that including somewhat larger matrix sizes will not allow one to decide on the issue. Although the large-deviation approach is rather powerful, even including $N=5000$ would require  substantial numerical resources on high-performance computers, not to speak of even larger sizes.

In the figure we also show, as a benchmark, the rate function obtained when the energy levels are independent, in the limit $N\to\infty$. In that case the rate function, together with its finite-size corrections, can be computed fully analytically (see \ref{app:FCS-independent}). For independent levels one finds that, upon taking an energy window of the size of the Thouless energy ($a /\sqrt{N}$ in this case) around zero, the rate function expressed in the appropriately rescaled variable $k$ approaches a constant as $k \to 0$, with the constant determined precisely by the window width $a$. 
In this simple case, one can also work out the finite-size corrections explicitly, and they behave in a manner similar to what we observe for comparable parameter values in Fig.~\ref{fig:counting:rare}(b). This benchmark is particularly illuminating in the present context, as we argue below. Indeed, the rate function of model~\eqref{eq:RPdef} is expected to exhibit a phase transition when the energy window is set on the scale of the Thouless energy. 
To see this, consider computing $I_N$ on an energy window of size $N^{-\alpha}$. If $\alpha < \gamma - 1$ (i.e.~the window is larger than the Thouless energy), then the rescaled rate function is expected to converge to that of independent levels, since on that scale the levels become effectively uncorrelated in the large-$N$ limit, and therefore no divergence as $k \to 0$ is expected in the rescaled variables. 
Conversely, if $\alpha > \gamma - 1$ (i.e.~the window is smaller than the Thouless energy), then the energy levels are strongly correlated and repel one another: the probability of finding a gap should be significantly smaller than for independent levels once the appropriate rescaling is applied, and the behavior as $k \to 0$ is expected to be qualitatively different --- possibly divergent --- since the independent-level rate function provides only a lower bound on $\Phi(k)$. This suggests that $\alpha = \gamma - 1$, which is precisely the value considered in Fig.~\ref{fig:counting:rare}, lies, most interestingly, exactly at the phase transition between these two regimes. This argument could therefore explain why the finite-size behavior observed here is so slow. 

On the other hand, we cannot exclude that
for large-enough values of $k$ a change of the behavior
appears.  Note that here the finite-size dependence of the empirical rate functions is very weak, and the results for large values of $N$ basically cannot be distinguished. Such changes of the rate function have been observed for the large-deviation properties of other models as well, e.g.~the height distribution of the Kardar--Parisi--Zhang equation~\cite{kpz2018}. 

\subsection{Level compressibility in the 
orthogonally invariant class
\label{sec:numerics-free}}

\begin{figure}[t]
    \centering
    \includegraphics[width=0.4\linewidth]{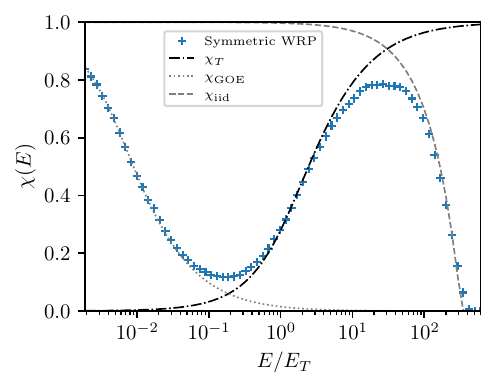}
    \hspace{20pt}
    \includegraphics[width=0.4\linewidth]{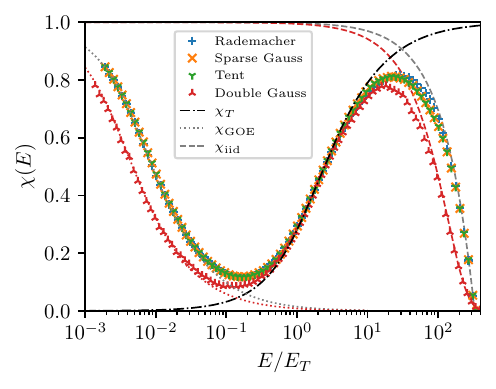}
    \put(-390,10){(a)}
    \put(-190,10){(b)}
    \caption{Level compressibility $\chi(E)$ for (a)~the Symmetrized Wishart--RP model, and (b)~for Wigner--RP matrices. 
    Symbols denote data from exact numerical diagonalization of random sample matrices of size $N=30000$. 
    For energies of the scales of the Thouless energy $E_T$~\eqref{eq:thouless}, all curves collapse on the universal prediction $\chi(E) = \chi_T(E/E_T)$ given in \cref{eq:chi_universal} and drawn with a dash-dotted line. 
    \rev{For low energy, the level compressibility is expected to approach the GOE prediction $\chi(E) = \chi_{\rm GOE}(E/\delta_N)$ in Eq.~\eqref{eq:chi_GOE}, indicated by dotted lines, where $\delta_N=1/(Np_a(0))$ is the mean level spacing. 
    By contrast, for $E\sim \order{1}$, the level compressibility approaches the prediction~\eqref{eq:chi-independent}, corresponding to the case of independent energy levels, indicated in the plot by dashed lines --- which is $p_a$-dependent.
    In panel~(b),}
    the first three sets of symbols correspond to the choice $p_a(a)=1/(2\sqrt{3})$, whereas the distribution $p_\zeta(x)$ is either Rademacher, sparse Gaussian, or tent, as described in \cref{sec:numerics-wigner}. 
    The red symbols in (b) correspond \rev{instead} to the case in which both $p_a(a)$ and $p_\zeta(x)$ are chosen to be the standard Gaussian distribution \rev{--- note that the predictions $\chi_\mathrm{GOE}$ and $\chi_\mathrm{iid}$ have to vary accordingly, and they are indicated in the plot by dotted or dashed red lines, respectively}.
    }
    \label{fig:LC_numerics}
\end{figure}

We now focus on the level compressibility, which we want to inspect for both classes of random matrices $\textbf{M}$ delineated in \cref{sec:generalizations}. First,
as an example of a random matrix with rotationally invariant statistical weight, which is at the same time relatively simple to generate and diagonalize,
we consider the case in which
$\newb$ in Eq.~\eqref{eq:RPdef} is a ``symmetrized'' Wishart matrix:
\begin{equation}
    \newb = \frac{1}{M}\left ( \textbf{W}_1 \textbf{W}_1^T - \textbf{W}_2 \textbf{W}_2^T \right)
    \; , 
    \label{eq:double-Wishart}
\end{equation}
where $\textbf{W}_{1,2}$ are two independent $N\times M$ matrices, with i.i.d.~random entries 
extracted from the normal distribution $\mathcal{N}(0, 1)$. 
With this choice, one can check that $\textbf{A}$ and $\newb$ in Eq.~\eqref{eq:RPdef} are indeed mutually free:
for any $\textbf{O} \in O(N)$, it is $ \textbf{W}_{i} \textbf{W}_{i}^T \overset{(law)}{=} \textbf{O}\textbf{W}_i \textbf{W}_i^T \textbf{O}^T $ for $i = 1,2$, hence
\begin{eqnarray}
    \left ( \textbf{W}_1 \textbf{W}_1^T - \textbf{W}_2 \textbf{W}_2^T \right) &\overset{(law)}{=} & \textbf{O}  \textbf{W}_1 \textbf{W}_1^T\textbf{O} ^T - \textbf{O}  \textbf{W}_2 \textbf{W}_2^T \textbf{O}^T 
    \nonumber\\
    &=& \textbf{O} \left ( \textbf{W}_1 \textbf{W}_1^T - \textbf{W}_2 \textbf{W}_2^T \right) \, \textbf{O}^T .
\end{eqnarray}
Moreover, we expect $\newb$ to be still quantitatively different from a GOE matrix, which formally corresponds to the sum of infinitely many free matrices $\textbf{W}_i$~\cite{potters2020first, Mingo2017}.

Numerical data for the corresponding level compressibility are shown in \cref{fig:LC_numerics}(a). We chose
$N = 30000$ and $N/M = 1/2$, with $p_a(a)$ uniform in $[-\sqrt{3}, \sqrt{3}]$, and we set the other parameters to $\nu=1$, $\gamma = 1.5$. We diagonalized $300$ samples and performed a moving average along the eigenvalue spectrum (which is determined by $p_a(a)$ in this regime, hence it is uniform).

The plot in Fig.~\ref{fig:LC_numerics}(a)
shows excellent agreement with the analytical prediction~\eqref{eq:chi_universal} for energies of the order of $E_T$, represented with a dash-dotted line.
As expected, for energies of the order of the mean level spacing $\delta_N=1/(Np_a(0))$, the level compressibility adheres instead to the GOE prediction $\chi(E) = \chi_{\rm GOE}(E/\delta_N)$, represented with a dotted line. Here (see e.g.~Appendix~E.3 in~Ref.\cite{Venturelli_2023})
    \begin{align} 
    \label{eq:chi_GOE} 
        \chi_{\rm GOE}(y) = & \, \frac{1}{2 \pi^2 y} \big\lbrace [\text{Si}(2 \pi y)]^2 -2\, \text{Ci}(4 \pi y) -\pi \, \text{Si}(2 \pi y)\nonumber\\
        & 
        \qquad
        +2 \left[-4 \pi y \, \text{Si}(4 \pi y)+2 \pi ^2 y+\ln (4 \pi y)-\cos (4 \pi y)+\gamma_E +1\right]\big\rbrace \, , 
    \end{align} 
    where ${\rm Ci}(z) = - \int_z^\infty \dd t \, \cos(t)/t $ and ${\rm Si}(z) = \int_0^z \dd t \, \sin(t)/t $ 
    are the cosine-integral and sine-integral functions, respectively, while $\gamma_E$ is the Euler–Mascheroni constant. 
Finally, for energies $E\sim \order{1}$, the numerical data are expected to deviate from $\chi_T$ and behave instead as in the case of independent energy levels~\cite{Venturelli_2023},
\begin{equation}
    \chi_{\rm iid}(E)=1-\int_{-E}^E \dd{\lambda} \, p_a(\lambda)
    \; ,
    \label{eq:chi-independent}
\end{equation}
and this is represented with a dashed line in Fig.~\ref{fig:LC_numerics}(a).

\subsection{Level compressibility in the Wigner class}
\label{sec:numerics-wigner}

Next, we test the universality of the level compressibility 
in the case in which 
$\newb$ is a Wigner matrix.
 To this end, we consider three choices of the distribution $p_\zeta(x)$ 
 of the 
 random variables $\zeta\ij$ in~\cref{eq:def_entries}:
\begin{enumerate}[(i)]
    \item Rademacher (bimodal): $\zeta_{ij}=0$ or $1$ with probability $1/2$.
    \item Sparse Gaussian: if $B_p$ is a Bernoulli variable (i.e.~non-zero with probability $0<p<1$), and $\mathcal N(0,1)$ a standard Gaussian variable, then $\zeta_{ij}=\mathcal N(0,1) \, B_p/\sqrt{p} $. Here we chose $p=0.1$ (in any case, one must assume $p\sim \mathcal{O}(N^0)$ for the saddle-point construction to hold\footnote{Note that the term
    \textit{sparse} is sometimes used in the literature to mean $p\sim 1/N$, which is however not the case analyzed here.}).
    \item ``Tent'', or triangular: $p_\zeta(x) = \frac{1}{\sqrt{6}}\left(1-\frac{|x|}{\sqrt{6}}\right)$.
\end{enumerate}

We have produced 
1000 instances 
of the matrix ${\bf M}$ in Eq.~\eqref{eq:def_entries}, of size $N=30000$. We then constructed ${\bf H}$ as in Eq.~\eqref{eq:RPdef}, with the elements of ${\bf A}$ chosen to be uniform in the interval $[-\sqrt{3},\sqrt{3}]$, hence with unit variance.
As in \cref{sec:numerics-free}, we chose $\nu=1$ and $\gamma=1.5$, and we
computed the level compressibility using a sliding average to improve the statistics.
The result is shown in Fig.~\ref{fig:LC_numerics}(b), and the agreement with the scaling function $\chi_T(y)$ is again remarkable in the region $E\sim E_T$.
Note that the three curves, corresponding to distinct choices of $p_\zeta(x)$, actually collapse not only in the region $E\sim E_T$, but on the entire spectrum --- which is expected here since $p_a(a)$ is the same in all three cases, but which is not a general feature. For comparison, we have thus added in the plot a curve corresponding to the case in which both $p_a(a)$ and $p_\zeta(x)$ are chosen to be standard Gaussians \rev{(red symbols)}, for which the expected collapse only happens around $E_T$ (no sliding average is possible in this case because the average spectral density is not uniform, so we averaged over a larger number of samples, namely $4000$).

\subsection{Discussion on the universality of the level compressibility}
\label{sec:discussion}
In the previous Sections, we have determined the CGF of the number of eigenvalues in an interval $[-E,E]$, 
given by \cref{eq:scaled-CGF,eq_scaling-function-G,eq:CGF-E-O-1,eq:r-theta}. This function describes the level statistics of 
GRP models in the form~\eqref{eq:RPdef}, where the matrix $\textbf{A}$ is diagonal with i.i.d.~entries, while the matrix $\newb$ can either be a Wigner matrix, or any random matrix mutually free with respect to $\textbf{A}$.
As a byproduct, we obtained that the level compressibility is given, for $E\sim E_T$, by the expression $\chi_T(E/E_T)$ 
in \cref{eq:chi_universal}.

In fact, the analytic form of $\chi_T$ was first derived in Ref.~\cite{Venturelli_2023} in the context of the standard (GOE/GUE) RP model, where it was noted that it does not depend on the choice of the distribution $p_a(a)$ of the entries of $\textbf{A}$. Its conjectured universality was later checked to hold for the Wishart--RP model~\cite{Delapalme2026}, where the matrix $\newb$ is drawn from the Wishart--Laguerre ensemble. This can be rationalized in the light of the results of \cref{sec:scaling-function}, since both these examples 
belong to the orthogonally invariant class which we analyzed in this work (see \cref{sec:free}).

Remarkably, the same scaling function $\chi_T$ turned out to describe the level compressibility also in the case in which the entries of $\newb$ are Lévy distributed~\cite{safonova2025}. This is not obvious from the analysis conducted in this work, because in \cref{sec:wigner} we assumed the variance $\cum_2$ of the entries of $\newb$ to be finite.

Assessing how much further this universality actually extends is a task that we will inevitably leave partially unfulfilled here.
In future work, we will address the impact of including correlations between the matrix elements of $\textbf{A}$, which is relevant in view of the modeling of more realistic interacting quantum systems~\cite{roy2024fock, roy2020localization, scoquart2024role,logan2025multifractality,Altshuler_2023}.
In the next Section, we take a different approach and we use numerics to directly measure the level compressibility in a simple but more realistic interacting system: the QREM~\cite{Faoro_2019,baldwin2018quantum,parolini2020multifractal,biroli2021out,kechedzhi2018efficient,Smelyanskiy_2020}.

\section{The quantum random energy model}
\label{sec:QREM}

In the previous Sections, we showed that the full counting statistics exhibits a universal scaling form across the crossover from random-matrix correlations at small energy scales to Poisson statistics at large energy scales. Remarkably, this scaling function is largely independent of the specific distributions of the matrices $\textbf{A}$ and $\newb$ entering the GRP models defined in Eq.~\eqref{eq:RPdef}.

A natural question is therefore to what extent this universality extends beyond random matrix ensembles and survives in realistic many-body quantum systems. To address this issue we consider the Quantum Random Energy Model (QREM)~\cite{Faoro_2019,baldwin2018quantum,parolini2020multifractal,biroli2021out,kechedzhi2018efficient,Smelyanskiy_2020}, one of the simplest models displaying many-body localization. This choice is motivated by the fact that a
mapping between the QREM and the RP model was previously proposed in Ref.~\cite{biroli2021out}, and successfully used to derive a qualitative description of its dynamical phase diagram, as we recall below.

The QREM for $\alpha=1, \dots, N_\sigma$ spin-$1/2$ degrees of freedom is defined by the Hamiltonian 
\begin{equation}
\hat{H}_{\rm QREM}
=
\hat E(\{\hat\sigma^z_\alpha\})
+
\Gamma\sum_{\alpha=1}^{N_\sigma}\hat\sigma^x_\alpha 
\; ,
\label{eq:qrem_hamiltonian}
\end{equation}
where $\Gamma$ denotes the transverse field, and
$\hat{E}(\{\hat\sigma^z_\alpha\})$ is a random operator diagonal in the computational ($z$-)basis. 
This operator assigns $2^{N_\sigma}$ uncorrelated energy values to the spin configurations, distributed according to the Gaussian density:
\begin{equation}
p(E)
=
\frac{\ee^{-E^2/N_\sigma}}{\sqrt{\pi N_\sigma}} 
\; .
\label{eq:gaussian_distribution}
\end{equation}
With this normalization the spectrum is concentrated, with overwhelming probability, in the interval
$[-N_\sigma\sqrt{\ln 2},N_\sigma\sqrt{\ln 2}]$
in the thermodynamic limit.
In the following, we denote by
$\varepsilon = E/N_\sigma$
the intensive energy density. 
The QREM arises as the $p\to\infty$ limit of the quantum $p$-spin Ising model, in which all groups of $p$
spins interact simultaneously through random couplings drawn from a Gaussian distribution~\cite{Goldschmidt90,NiRi98,CuGrDa00,CuGrDa01}.
The diagonal part of the Hamiltonian coincides with the classical Random Energy Model, a paradigmatic mean-field spin-glass model~\cite{derrida1980random}.

The QREM can be interpreted as an Anderson localization problem in Hilbert space. 
Indeed,
choosing the simultaneous eigenstates of all operators $\hat \sigma^z_\alpha$ as a basis, each spin configuration corresponds to a vertex of an $N_\sigma$-dimensional hypercube containing
$N=2^{N_\sigma}$
vertices.
The random energies provide uncorrelated on-site disorder, while the transverse-field term induces hopping between nearest-neighbor vertices differing by a single spin flip.
The many-body Hamiltonian can therefore be rewritten as
\begin{equation}
\textbf{H}_{\rm QREM}
=
\sum_{i=1}^{N} E_i |i\rangle\langle i|
+
\Gamma
\sum_{\langle i,j\rangle}
\left(
|i\rangle\langle j|
+
|j\rangle\langle i|
\right),
\label{eq:anderson_hypercube}
\end{equation}
where $\langle i,j\rangle$ denotes nearest-neighbor vertices of the ${N_\sigma}$-dimensional hypercube of $N=2^{N_\sigma}$ vertices $\{+1,-1\}^{N_\sigma}$.
This mapping is exact: Eqs.~\eqref{eq:qrem_hamiltonian} and~\eqref{eq:anderson_hypercube} describe the same spectrum and eigenstates when expressed in the basis of the $\{ \hat{\sigma}_a^z \}$.

\subsection{Effective RP description}

The QREM Hamiltonian~\eqref{eq:anderson_hypercube} 
formally takes the form
\(
\mathbf H_{\rm QREM}
=
\mathbf A+\mathbf M
\),
where
\(
\mathbf A
=
\mathrm{diag}(E_1,\ldots,E_N)
\)
and
\(
\mathbf M
\)
is the adjacency matrix of the hypercube multiplied by $\Gamma$.
Unlike the GRP ensembles considered in Section~\ref{sec:scaling-function} and
Section~\ref{sec:numerics}, 
here $\mathbf M$ is neither orthogonally invariant nor a Wigner matrix. Consequently, 
the universality established in these Sections is not \emph{a priori} 
expected  to hold here.

Crucially, while the hopping $\Gamma$ does not scale with $N$, the role of the scaling parameter in the GRP models of Eq.~\eqref{eq:RPdef} is effectively played by the intensive energy $\varepsilon$. 
Indeed, in the small-$\Gamma$ limit, the density of states is dominated by the diagonal part, $\rho(\varepsilon) \simeq p(\varepsilon) \sim \ee^{-{N_\sigma}\varepsilon^2}$, leading to a mean level spacing:
\begin{equation}
\delta_{N_\sigma}(\varepsilon)
=
\frac{1}{2^{N_\sigma}\rho(\varepsilon)}
\simeq
\ee^{\,{N_\sigma}(\varepsilon^2-\ln2)} .
\label{eq:DoS}
\end{equation}
Since $\delta_{N_\sigma}(\varepsilon)$ depends exponentially on $\varepsilon$, the energy density allows one to parametrically tune the ratio between the disorder and the hopping. In this sense the parameter $\varepsilon$ plays a role analogous to the control parameter $\sqrt{\etanew}$ of the GRP model~\eqref{eq:RPdef}: by varying $\varepsilon$, one changes the ratio between the characteristic diagonal level spacing and the off-diagonal hybridization scale. 

Based on these ideas, following Ref.~\cite{biroli2021out}, one can derive an effective RP description of the QREM using a forward-scattering approximation~\cite{anderson1958absence}. Consider two spin configurations $|0\rangle$ and $|f\rangle$ of energy $E_{|0\rangle} \simeq E_{|f\rangle} \simeq E$ separated by a Hamming distance ${N_\sigma}x$, with $0<x<1$. 
(The latter is defined as the minimum number of spin flips separating the two configurations on the Hilbert space graph.) Since the diagonal energies are independent, the typical number of configurations at energy density $\varepsilon$ and distance ${N_\sigma}x$ from a given configuration is
\begin{equation}
\mathcal N_\varepsilon(x)
=
\binom{{N_\sigma}}{{N_\sigma}x}
\frac{\ee^{-{N_\sigma}\varepsilon^2}}{\sqrt{\pi {N_\sigma}}} .
\label{eq:number_configurations}
\end{equation}
Within the forward-scattering approximation~\cite{anderson1958absence,ros2015integrals,imbrie2017local,pietracaprina2016forward,tarzia2020many,tarzia2022fully}, the matrix element connecting two such configurations is obtained by summing the contributions of all shortest paths between them, while neglecting higher-order loop corrections in which spins are flipped twice, since they contribute at higher order in perturbation theory:
\begin{equation}
{\cal B}_{|0\rangle \to |f\rangle} \approx \sum_{{\cal P} \in \{ \text{shortest paths~}|0\rangle \to |f\rangle \} } \, \prod_{|m\rangle \in {\cal P}} \frac{\Gamma}{E - E_{|m\rangle}} \, .
\end{equation}
Almost all states $|m\rangle$ crossed along these paths have energy $E_{|m\rangle} \sim \order {\sqrt{{N_\sigma}}}$, while the energy of the initial and final configurations are  $E\approx \order{{N_\sigma}}$, therefore the energy differences appearing in the denominators of the perturbative expansion can be replaced by $E={N_\sigma} \varepsilon$. Since there are $({N_\sigma}x)!$ shortest paths of length ${N_\sigma}x$ between the two configurations, corresponding to the $({N_\sigma}x)!$ ways to connect two configurations that are ${N_\sigma}x$ spin flips away (resulting from the permutations of the order of the spins that are flipped), the resulting effective tunneling amplitudes read
\begin{equation}
{\cal B}(\varepsilon,x)
\simeq 
\left(
\frac{\Gamma}{{N_\sigma}\varepsilon}
\right)^{{N_\sigma}x}
({N_\sigma}x)! \;  .
\label{eq:matrix_element}
\end{equation}
These tunneling amplitudes play the role of the matrix elements $m_{ij}$ of the GRP models defined in Eq.~\eqref{eq:RPdef}. 

\begin{figure}
    \centering
    \includegraphics[width=0.5\linewidth]{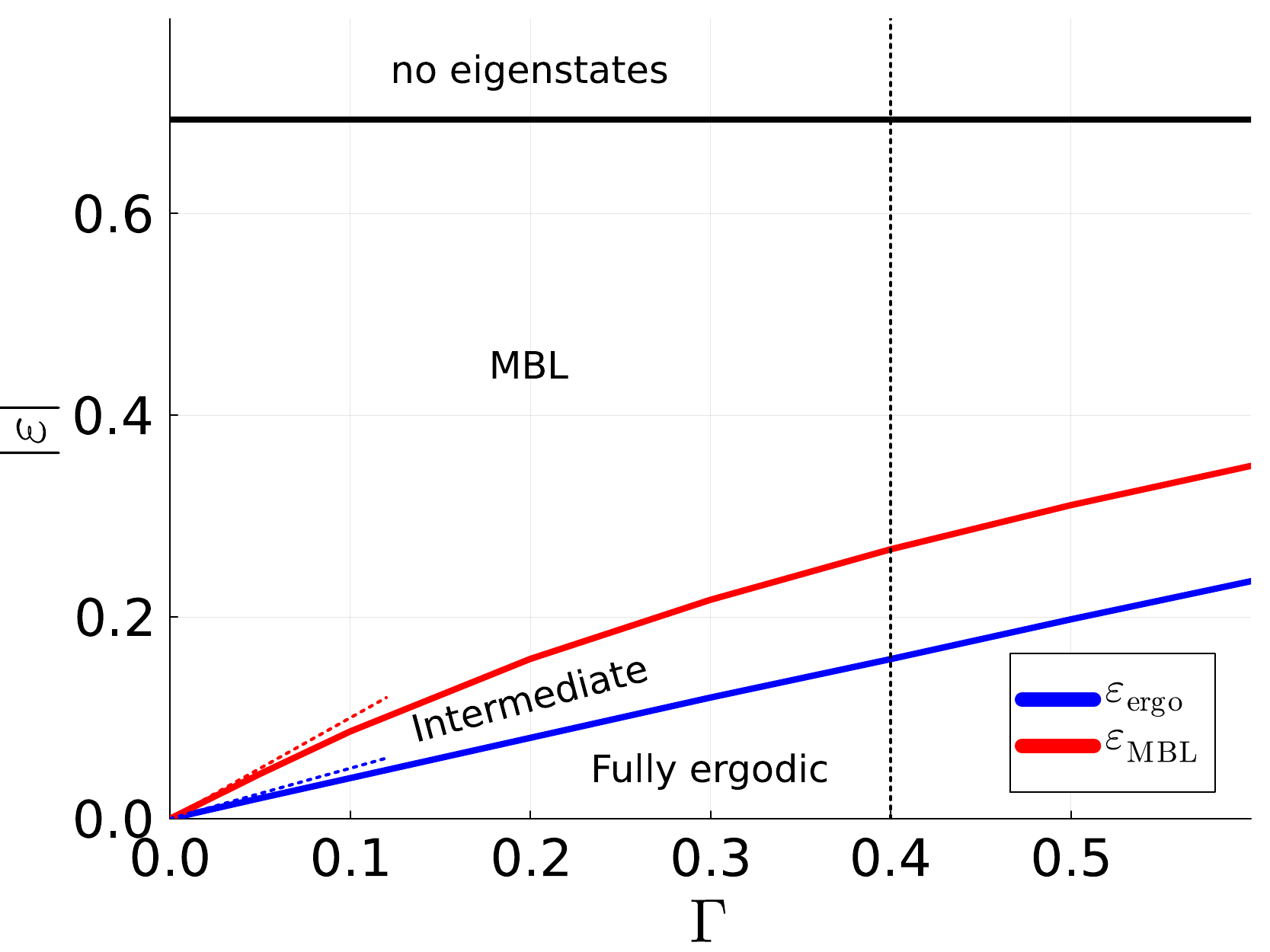}
    \caption{Schematic out-of-equilibrium phase diagram of the QREM (see also Refs.~\cite{Faoro_2019,baldwin2018quantum,parolini2020multifractal,biroli2021out,kechedzhi2018efficient,Smelyanskiy_2020}). The transition line separating the fully ergodic phase from the intermediate partially delocalized, non-ergodic phase (blue), and the transition line separating the intermediate phase from the localized phase (red), are obtained from the numerical solution of Eqs.~\eqref{eq:gamma_x}, \eqref{eq:transition_condition_1}, and \eqref{eq:transition_condition_2} (see Ref.~\cite{biroli2021out} for details). The dashed lines correspond to the linear asymptotic behavior of the transition lines, Eq.~\eqref{eq:expansion}, obtained from a small-$\Gamma$ expansion. The horizontal black line at $|\varepsilon|=\ln 2$ shows the position of the edge of the many-body spectrum beyond which no states are found. 
    Since the spectrum is statistically symmetric with respect to $\varepsilon=0$, we only show the upper half-plane. The vertical  dotted line marks the value of the transverse field, $\Gamma=0.4$, at which all numerical results discussed in this section have been obtained. 
    }
    \label{fig:pdQREM}
\end{figure}

Comparing the scaling of the effective matrix elements with the number of spin configurations $\mathcal{N}_{\varepsilon}(x)$ at energy $\varepsilon$ and Hamming distance ${N_\sigma}x$ from the initial configuration suggests to introduce an effective RP exponent $\gamma_x$ through
\begin{equation}
{\cal B}(\varepsilon,x)
\propto
\left[\mathcal N_\varepsilon(x)\right]^{-\gamma_x/2}.
\end{equation}
Using Stirling's approximation at leading order one obtains
\begin{equation} 
\gamma_x
\simeq
\frac{
2x\ln\!\left(\frac{\varepsilon \ee}{\Gamma x}\right)
}
{
\ln 2
-
\varepsilon^2
+
x\ln x
+
(1-x)\ln(1-x)
} \, .
\label{eq:gamma_x}
\end{equation}
The effective RP description suggests that delocalization occurs whenever at least one Hamming-distance $x$-sector becomes delocalized. Following the RP phenomenology, this corresponds to the condition
\begin{equation} 
\min_x \gamma_x = 2 \, ,
\label{eq:transition_condition_1}
\end{equation}
which defines the transition between localized and non-ergodic extended states.

Similarly, full ergodicity is expected to be restored when at least one $x$-sector enters the fully ergodic RP regime,
\begin{equation} 
\min_x \gamma_x = 1 \, .
\label{eq:transition_condition_2}
\end{equation}
The resulting phase diagram consists of three distinct regions~\cite{anderson1958absence,ros2015integrals,imbrie2017local,pietracaprina2016forward,tarzia2020many,tarzia2022fully}, as represented in Fig.~\ref{fig:pdQREM}. At low energy density, the eigenstates are fully ergodic and spread over the whole accessible Hilbert space volume ${\cal N}(\varepsilon) = 2^{N_\sigma} \rho(\varepsilon)$ at energy $\varepsilon$. 
At large energy density, the system is localized and eigenstates of energy ${N_\sigma} \varepsilon$ remain confined to a small neighborhood of a given spin configuration of energy ${N_\sigma} \varepsilon$. Between these two regimes lies an intermediate partially delocalized but non-ergodic phase, in which eigenstates hybridize with an exponentially large yet subextensive number of configurations. 
In this regime, the fraction of Hilbert space occupied by an eigenstate vanishes in the thermodynamic limit. Expanding the solutions of Eqs.~\eqref{eq:transition_condition_1} and~\eqref{eq:transition_condition_2} around $\varepsilon=0$ yields
\begin{equation} \label{eq:expansion}
\Gamma_{\rm MBL}
\simeq
\varepsilon,
\qquad
\Gamma_{\rm erg}
\simeq
\frac{\varepsilon}{2},
\end{equation}
in agreement with the phase diagram proposed in Ref.~\cite{biroli2021out} (see also Refs.~\cite{Faoro_2019,baldwin2018quantum,parolini2020multifractal}).

\subsection{Level compressibility}

Although the effective RP description relies on several uncontrolled approximations --- including the forward-scattering approximation, the use of typical tunneling amplitudes, and the reduction of the problem to the most resonant Hamming-distance sector --- it provides a useful phenomenological framework for understanding the structure of QREM eigenstates. This naturally raises the question as
to  whether the universal scaling of the full counting statistics derived in Section~\ref{sec:scaling-function} also emerges in the non-ergodic extended phase of the QREM.

To address this issue, we computed numerically 
 the first two cumulants,
$\kappa_1(\omega)$ and $\kappa_2(\omega)$,
of the number of eigenvalues $I_N [{N_\sigma} \varepsilon - \omega, {N_\sigma} \varepsilon + \omega]$ contained in a symmetric energy window of width $\omega$ centered around the energy
$E={N_\sigma}\varepsilon$.
We generated matrices of size~$2^{N_{\sigma}}\times 2^{N_{\sigma}}$  according to \cref{eq:anderson_hypercube}, i.e.~with Gaussian random numbers drawn according to  \cref{eq:gaussian_distribution} on the diagonal, and entries $\Gamma>0$ off diagonal whenever the corresponding row and column indices (in binary form) correspond to neighbors on the $2^{N_\sigma}$-dimensional hypercube. We diagonalized these matrices numerically and counted the number of eigenvalues in the desired interval, as before.
Throughout this Section we focus on
$\Gamma=0.4$,
for which the effective RP description predicts an extended non-ergodic regime over a broad range of energy densities from $\varepsilon_{\rm erg} \approx 0.158$ to $\varepsilon_{\rm MBL} \approx 0.29$.

\begin{figure}
    \centering
    \includegraphics[width=0.335\linewidth]{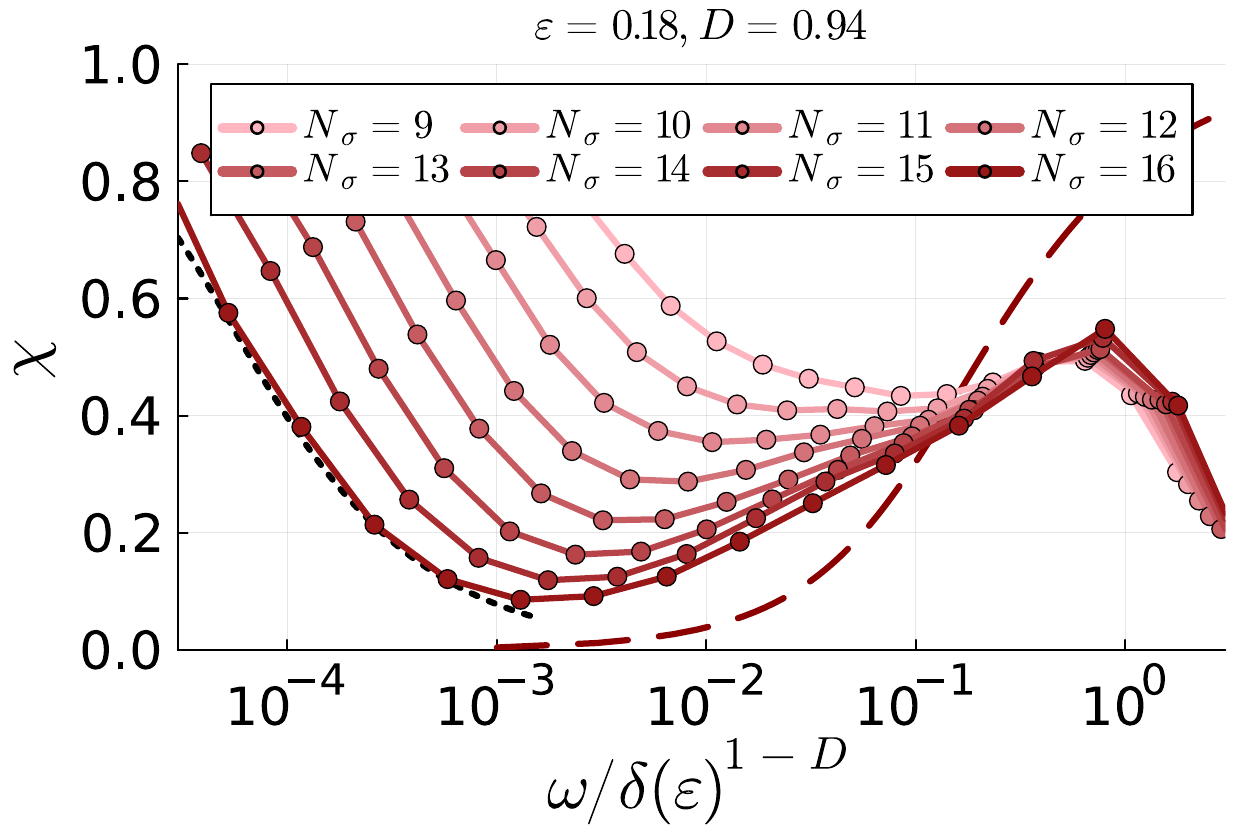}
    \hspace{-10pt}
    \includegraphics[width=0.335\linewidth]{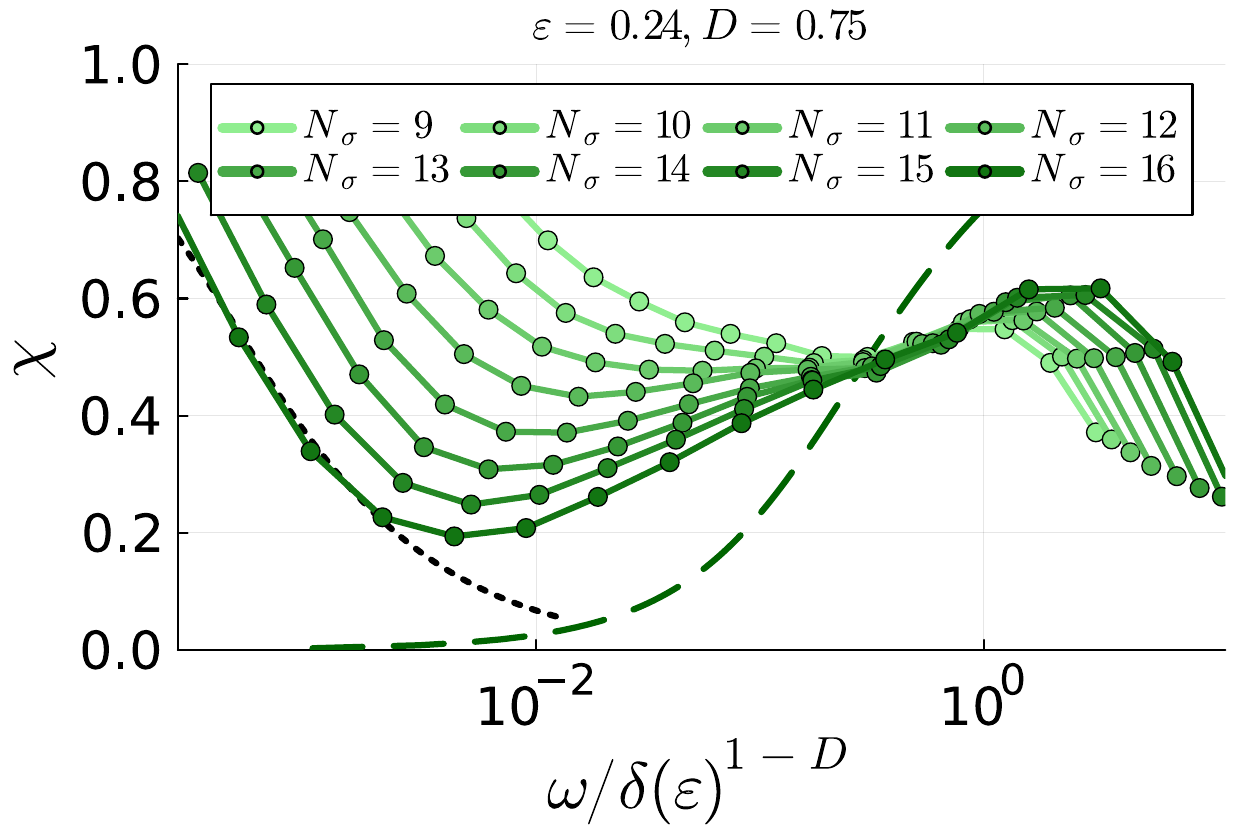}
    \hspace{-10pt}
    \includegraphics[width=0.335\linewidth]{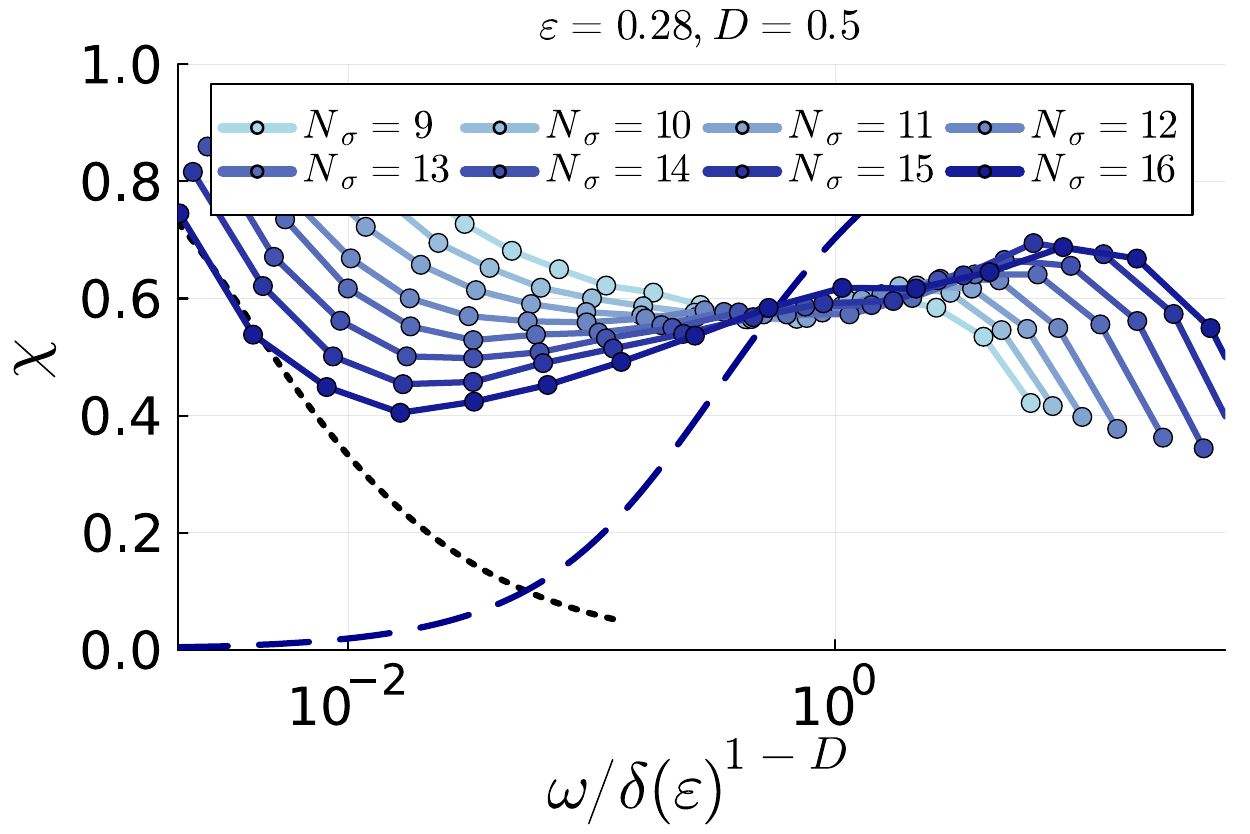}
    \put(-335,26){\tiny (a)}
    \put(-182,26){\tiny (b)}
    \put(-30,26){\tiny (c)}
    \caption{Level compressibility $\chi(\omega)$ for the QREM, Eq.~\eqref{eq:qrem_hamiltonian}, at $\Gamma=0.4$ for (a)~$\varepsilon=0.18$, (b)~$\varepsilon=0.24$, and (c)~$\varepsilon=0.28$, spanning the intermediate non-ergodic delocalized phase of the model (see Fig.~\ref{fig:pdQREM}). 
    The $\omega$-axis is rescaled by 
    $(\delta_{N_\sigma}(\varepsilon))^{1-D}$, with an $\varepsilon$-dependent exponent $D$ which produces the best collapse of the curves for different system sizes ${N_\sigma}$. The dashed curves represent the  crossover scaling function given in Eq.~\eqref{eq:chi_universal}.
    The black dotted curve at low energies corresponds to the universal GOE prediction~\eqref{eq:chi_GOE} (shown only for the largest available system size, $N_\sigma=16$), which accurately reproduces the behavior of the level compressibility on the scale of the spectral gap.}
    \label{fig:chiQREM}
\end{figure}

As in the GRP models studied in the previous Sections, we focus on the level compressibility,
\begin{equation}
\chi(\omega)
=
\frac{\kappa_2(\omega)}
{\kappa_1(\omega)}.
\end{equation}
Figure~\ref{fig:chiQREM} shows the level compressibility for three values of the energy across the intermediate phase and several  system sizes ${N_\sigma}$. To investigate the existence of a scaling regime, we rescale the energy window $\omega$ by a generalized Thouless energy,
\begin{equation}
E_{T}
\propto
\delta_{N_\sigma}(\varepsilon)
\,
{\cal N}(\varepsilon)^D
=
\delta_{N_\sigma}(\varepsilon)^{\,1-D}
\; ,
\end{equation}
where
${\cal N}(\varepsilon)=2^{N_\sigma}\rho(\varepsilon)$
is the number of Hilbert-space configurations at energy density $\varepsilon$, and $D$ is an effective fractal dimension.

Within the effective RP picture one would expect
$D$ to be related to the exponent $\gamma_x$ characterizing the most resonant sector $x_\star$ as $D = 2 - \gamma_{x_\star}$. More generally, however, we determine $D$ directly from the numerical data by requiring the best collapse of $\chi(\omega/E_{\rm T})$ on a single scaling function as the system size is varied. The mean level spacing $\delta_{N_\sigma}(\varepsilon)$ is measured directly from exact diagonalization.

The resulting data collapses found in Fig.~\ref{fig:chiQREM} indicate the existence of a well-defined scaling regime. However, the associated scaling function differs markedly from the universal crossover function~\eqref{eq:chi_universal} found for the GRP ensembles studied in the previous sections. In particular, the crossover between RMT behavior at small energy scales and Poisson statistics at large energy scales is substantially broader than predicted by Eq.~\eqref{eq:chi_universal}. Furthermore, the scaling function itself depends on the energy density $\varepsilon$. As the energy density decreases and the system approaches the ergodic phase, the crossover becomes progressively sharper and the scaling function moves closer to the RP prediction~\eqref{eq:chi_universal}. Conversely, upon approaching the localized phase, the crossover broadens significantly and departs more and more from the universal curve observed in random matrix models.

These results provide direct evidence that the universality established in Sections~\ref{sec:Path-integral}, \ref{sec:scaling-function} and 
\ref{sec:numerics} does not extend to the QREM. Although the model admits an effective RP interpretation at the level of the phase diagram~\cite{Facoetti_2016}, its spectral correlations retain non-trivial features that are not captured by the random matrix description.

\subsection{Discussion}
\label{sec:discussion-qrem}

A plausible explanation for the breakdown of universality is that the structure of the QREM eigenstates is considerably more complex than that of RP eigenvectors. Indeed, in the RP model the non-ergodic phase is characterized by a single \textit{fractal exponent} controlling the scaling of the support set (see \cref{sec:fractal}). 
Actually, as shown in full generality in~\cite{Kutlin_2024}, the eigenstates of all matrix ensembles of the form~\eqref{eq:RPdef}, as the ones considered in this paper, are also fractal as far as the elements of $\textbf{A}$ and $\textbf{M}$ are uncorrelated. 
By contrast, our numerical results suggest that the QREM eigenstates might exhibit \textit{multi}fractal properties: different moments of the wave-function amplitudes scale with distinct exponents (i.e.,~$D_q$ 
varies with $  q$), implying that no single fractal dimension can fully characterize the eigenstate statistics.

\begin{figure}
    \centering
    \includegraphics[width=0.5\linewidth]{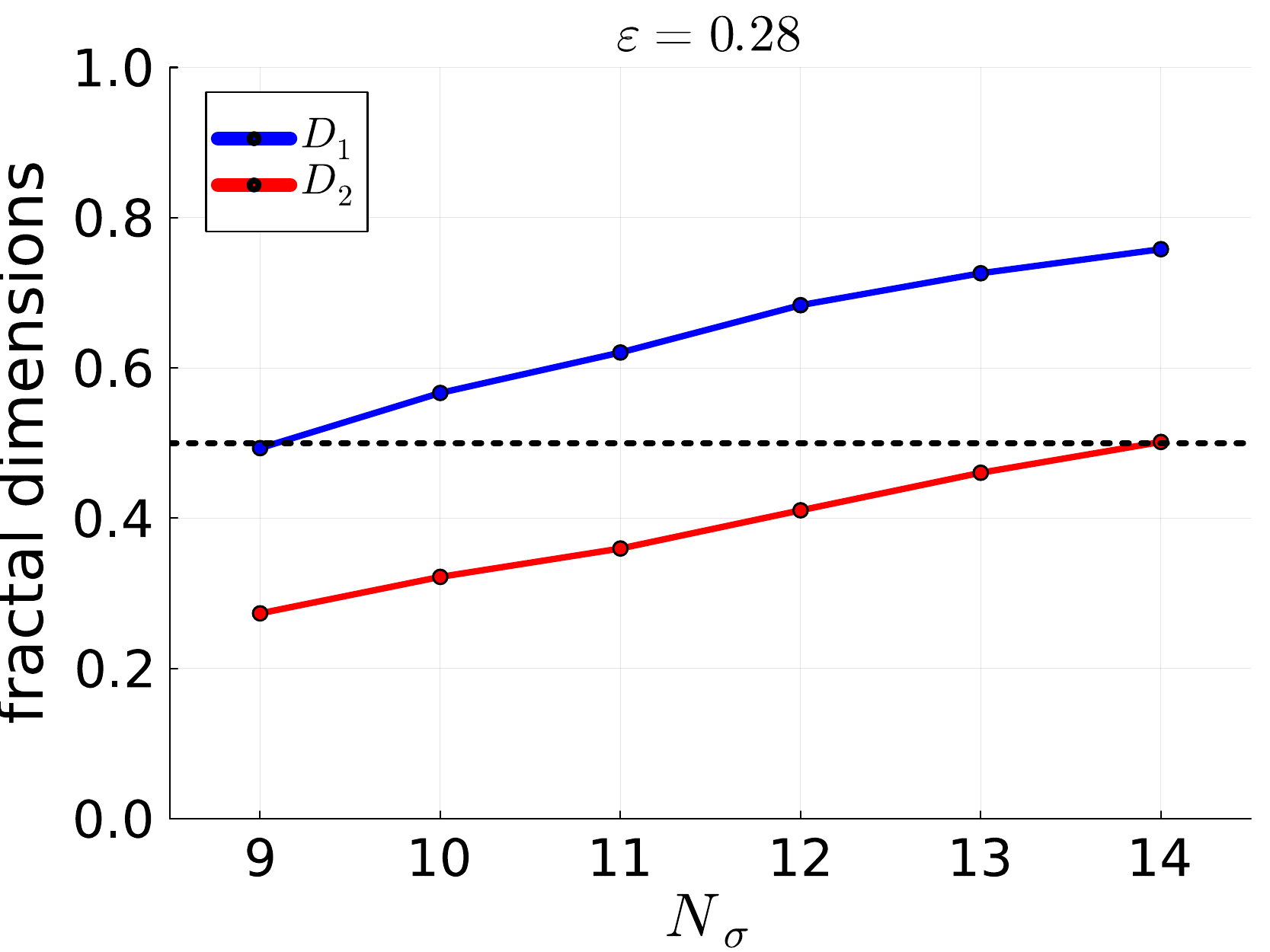}
    \caption{Fractal dimensions $D_1$ (blue) and $D_2$ (red) of the QREM eigenstates for $\Gamma=0.4$ and $|\varepsilon|=0.28$, computed numerically from Eq.~\eqref{eq:fractal}. The horizontal black dotted line indicates the value of the fractal exponent $D$ that yields the best collapse of the level compressibility data for different system sizes (see Fig.~\ref{fig:chiQREM}(c)).}
    \label{fig:fractal}
\end{figure}

To illustrate this point, Fig.~\ref{fig:fractal} compares the effective exponent $D$ extracted from the collapse of the level compressibility with the generalized (flowing) fractal dimensions $D_1$ and $D_2$ obtained independently from the scaling of the eigenstate 
moments as
\begin{equation} \label{eq:fractal}
\begin{aligned}
    D_2 & = - \ln \left \langle \frac{\sum_\alpha  \left( \sum_{i=1}^N \vert \langle i |\psi_\alpha \rangle \vert^{4} \right) 1_{E_\alpha \in [N_\sigma (\varepsilon - \Delta_\epsilon),N_\sigma (\varepsilon + \Delta_\epsilon)]}
    }{\sum_\alpha 1_{E_\alpha \in [N_\sigma (\varepsilon - \Delta_\epsilon),N_\sigma (\varepsilon + \Delta_\epsilon)]}
    } \right \rangle \Bigg / \ln {\cal N}(\varepsilon)\, , \\
    D_1 & = - \left \langle  \frac{\sum_\alpha  \left( \sum_{i=1}^N \vert \langle i |\psi_\alpha \rangle \vert^{2} \ln \vert \langle i |\psi_\alpha \rangle \vert^{2}   \right) 1_{E_\alpha \in [N_\sigma (\varepsilon - \Delta_\epsilon),N_\sigma (\varepsilon + \Delta_\epsilon)]}
    }{\sum_\alpha1_{E_\alpha \in [N_\sigma (\varepsilon - \Delta_\epsilon),N_\sigma (\varepsilon + \Delta_\epsilon)]}
    } \right \rangle \Bigg / \ln {\cal N}(\varepsilon) \, ,
    \end{aligned}
\end{equation}
where $1_{E_\alpha \in [N_\sigma(\varepsilon-\Delta_\varepsilon),\,N_\sigma(\varepsilon+\Delta_\varepsilon)]}$ is an indicator function selecting a small energy window of width $2\Delta_\varepsilon$ (with $\Delta_\varepsilon=0.01$), within which the average moments of the eigenstates are computed.
The figure shows that, within the numerical accuracy of our calculations and over the range of system sizes accessible to exact diagonalization, these three exponents are clearly distinct. This observation suggests that the mini-band structure governing spectral correlations in the QREM differs qualitatively from that of the RP ensemble, and cannot be characterized by a single scaling exponent. 
It also suggests that 
the universality discussed in Sections~\ref{sec:model}--\ref{sec:numerics} 
may be restricted to random-matrix ensembles whose eigenstates are characterized by a single fractal dimension, rather than by a fully multifractal spectrum.

Remarkably, however,
although the scaling function observed in the QREM differs from the universal one found for the RP-type ensembles, a well-defined scaling regime still emerges throughout the intermediate non-ergodic extended phase. 
This suggests that one may be able to suitably extend and generalize the random-matrix models studied in this work to capture some fundamental aspects of spectral statistics in the crossover regime between ergodicity and localization, even though the GRP models considered here fail to reproduce the full phenomenology quantitatively.

\section{Conclusions}
\label{sec:conclusion}

We have characterized the level statistics in various RP-type models, consisting of the sum of a diagonal matrix $\textbf{A}$ with entries independently distributed according to $p_a(a)$, and another random matrix $\textbf{M}$, as in \cref{eq:RPdef}. In the standard RP model, $\textbf{M}$ is a Gaussian random matrix, whereas here we explored the cases in which $\textbf{M}$ is either an orthogonally invariant matrix (hence mutually free with respect to $\textbf{A}$), or a Wigner matrix (with independent entries extracted from $p_\zeta(x)$, not necessarily Gaussian).

Combining tools from free probability theory and the replica method, we have computed the full counting statistics --- that is, the cumulant generating function~\eqref{eq:CGF-def} of the number $I_N[-E,E]$ of eigenvalues lying in the symmetric interval $[-E,E]$ --- up to leading order for large system size $N$, in the fractal phase $1<\gamma <2$ of the model. The result is reported in \cref{eq:CGF-E-O-1,eq:r-theta}; 
it depends explicitly on $p_a(a)$, whereas the particular choice of $\textbf{M}$ only enters through a single parameter, namely either the second free cumulant $\free_2$ in the orthogonally invariant case, or the variance $\cum_2$ of $p_\zeta(x)$ in the Wigner case \rev{(provided that $\free_2$ and $\cum_2$ are finite, respectively)}.

In the case in which the size of the interval is chosen to be of the order of the Thouless energy $E\sim E_T$~\eqref{eq:thouless}, such cumulant generating function takes a simple scaling form, \cref{eq:scaling-function-def,eq_scaling-function-G}, depending only on the ratio $E/E_T$, and universal across all variations of the model described above. As a by-product, one finds that the level compressibility~\eqref{eq:chi_universal} (obtained as the ratio of the first two cumulants of $I_N$) is also universal in this sense, thus confirming the conjecture originally put forward in~\cite{Venturelli_2023}.

We have supported our conclusions numerically in \cref{sec:numerics} by measuring both the level compressibility \rev{(Figure~\ref{fig:LC_numerics})} and the large-deviation function \rev{(Figure~\ref{fig:counting:rare})}. The former, studied for matrices with sizes of the order of $N \sim 10^4$,  
satisfies the universal scaling found analytically. 
The latter could only be sampled here for much smaller matrix sizes, $N\sim 10^3$, and suffers from strong finite-size effects. 
The rate function agrees well with the analytic prediction close to its minimum,
while larger deviations persist away from it; these appear to slowly decrease upon increasing $N$, although the convergence is too slow to conclusively establish agreement in this regime. 
Improving the numerical data by using larger sizes lies beyond our present numerical capabilities.

Finally, in \cref{sec:QREM} we have tested whether this universality extends to the QREM, which is one of the simplest interacting models displaying many-body localization. Our main finding is that, even within this setting, the level compressibility exhibits a collapse as a function of $E/E_T$ within the intermediate non-ergodic phase of the model 
\rev{(Figure~\ref{fig:chiQREM})}, suggesting that a matrix model may be able to capture the level statistics in this scaling regime. However, such scaling function does not agree quantitatively with the one found within the GRP model. We have imputed this difference to the fact that, contrary to the RP model, the eigenstates in the non-ergodic phase of the QREM are \textit{multi}fractal~\cite{Kutlin_2024}
and not simply fractal, thus giving rise to a more complex mini-band spectral structure (see \cref{sec:discussion-qrem} \rev{and Figure~\ref{fig:fractal}}).

These findings motivate further investigation of the level statistics in variations of the RP model which could display multifractality, such as random Cantor sets~\cite{Altshuler_2023}, Power Law Random Banded matrices~\cite{Mirlin96}, or in the Ultra metric model~\cite{Fyodorov_2009, swietek2026}.
Another important extension concerns the inclusion of correlations between the matrix elements of $\textbf{A}$~\cite{Altshuler_2023}, so as to mimic more closely the correlations between energy levels that are inevitably present in more realistic quantum systems.
Finally, recent developments~\cite{craps2026} pointed out that heavy-tailed hopping amplitudes (as in the Lévy-RP model~\cite{biroli2021levy}) appear to be generic in the integrability-to-chaos crossover of many quantum systems, thus motivating a deeper analysis of toy random matrix models including this feature.

 \section*{Acknowledgments}
We thank Oleg Evnin and Guilhem Semerjian for interesting discussions.
DV thanks Pascal Viot for his help with numerical resources.
This work was conducted in the spirit of the Slow Science Manifesto, advocating for collaborative and sustainable research (\href{https://www.slow-science.com}{slow-science.com}). 
We acknowledge financial support from the ANR research grant ManyBodyNet ANR-24-CE30-5851. AKH thanks the Centre National de Recherche (CNRS) for financial support, and the LTPMC and LPTHE for their hospitality. Some of the simulations were performed on the
    HPC cluster ROSA, located at the University of Oldenburg (Germany), and
    funded by the DFG through its Major Research Instrumentation Program
    (INST 184/225-1 FUGG) and the Ministry of Science and Culture (MWK) of the
    Lower Saxony State.
 
\appendix
\addtocontents{toc}{\fixappendix}

\section{\rev{Generalization to Hermitian matrices}}
\label{app:complex}

\rev{As we mentioned in the footnote~\ref{foot:unitary}, the free probability approach can also be applied to complex Hermitian matrices 
that are \textit{unitarily} invariant ---
i.e.~matrices whose 
probability law is invariant under the action of a unitary matrix, $P[ \textbf{M}  ] =  P [\textbf{UMU}^T]$ for any $\textbf{U} \in U(N)$. 
In this setting, the integrals with respect to the Haar measure of the $O(N)$ group 
in Section~\ref{sec:free} have to be replaced by
integrals over the $U(N)$ group. It is again possible to use the low rank HCIZ integral
in the form~\cite{potters2020first}\begin{equation}
    I ( \textbf{X}, \textbf{Y}) \equiv \Biggl \langle 
    \ee^{
    \frac{N}{2} \Tr 
    [\textbf{UXU}^T \textbf{ Y} ]
    }
    \Biggr \rangle_{\textbf{U}} \approx \exp \left ( \frac{\beta N}{2} \Tr \mathcal{H}_{\textbf{X}}(\textbf{Y}) \right ),
\end{equation} 
where $\beta$ is the Dyson index, which is equal to 2 if $\textbf{U} \in U(N)$.
The calculation then proceeds in the same way as in Section~\ref{sec:free},
and it yields an additional factor $\beta = 2$ in front of the second line of Eq.~\eqref{eq:action}. 
Upon setting $ \free _1 = 0$, and expanding up to $\mathcal O (\etanew)$ 
while assuming 
$\free _2$ to be finite,
we get that Eq.~\eqref{eq:action_order_eta_2} 
generalizes in this case to
\begin{align}
    \mathcal{S}_n [\mu, \tilde \mu ; \underline \Lambda] = & -{\rm i} \int \mathrm{d} \vec{y} \, \mu (\vec{y}) \, \tilde \mu (\vec{y}) + \ln \int \mathrm{d} \vec{y} \, \ee^{-\frac{{\rm i}}{2} \vec{y}^{\, T} \underline \Lambda \vec{y} + {\rm i} \tilde \mu ( \vec{y})} \: \psi_a \left ( - \frac{1}{2} \vec{y}^{\; T} \underline L  \vec{y} \right ) 
    \nonumber\\
    &  - \beta \frac{\etanew \free _2}{4}\int \mathrm{d} \vec{y} \,\mathrm{d} \vec{ w } \,  \mu (\vec{y}) \, \mu (\vec{w}) \, \left ( \vec y ^{\, T}  \underline L \vec w \right )^2 + o (\etanew )
    \; .
    \label{eq:action-beta}
\end{align} 
Therefore, up to 
an additional
factor $\beta$
which renormalizes $\free _2$, we conclude that
the unitarily invariant ensemble 
shares the same full counting statistics
as its real symmetric counterpart (i.e.~the orthogonally invariant ensemble with the same free cumulants).
Since this is true in principle for all energies, we deduce that
the universality of the full counting statistics around the Thouless energy $E_T$ also holds for 
unitarily invariant matrices. 
This result is not surprising, 
given that the distributions of the real eigenvalues of the real and complex Gaussian ensembles are in fact very similar.}

\rev{We can 
similarly
extend the Wigner case
considered in Section~\ref{sec:wigner}
to Hermitian matrices, by replacing $ m_{ij} $ by $ m_{ij} + i \hat m_{ij} $ in Eq.~\eqref{eq:def_M_wigner}, where the $\hat m_{ij}$ are also i.i.d.~random variables sampled from 
the same distribution $p_{\zeta}$ as the corresponding real parts  $ m_{ij} $. 
Following similar steps, one can check that
Eq.~\eqref{eq:M-def}
generalizes in this case to 
\begin{equation}
    \tilde{ \mathcal{K}}(\vec{y}, \vec{w}) \equiv N \,F\left( \sqrt{\frac{\etanew}{2N}} \vec{y}^{\; T} \underline L \vec{w} \right) + N \,F\left( \ii \sqrt{\frac{\etanew}{2N}}\vec{y}^{\; T} \underline L \vec{w} \right).
\end{equation}
Upon focusing on the symmetric case by
setting $\cum _1 = 0$, and expanding 
up to $\mathcal{O}(\etanew)$,
\cref{eq:def_M_wigner_lead}
becomes $ \tilde {\mathcal K}(\vec{y}, \vec{w}) = - \beta \frac{\etanew \cum _2}{4} (\vec{y}^{\; T} \underline L \vec{w})^2 + o (\etanew) $ with $\beta = 2$, 
and the resulting action thus matches the one in \cref{eq:action-beta}.
Again, this
proves that universality 
for energies on the scale of 
$E_T$
also holds for complex Hermitian Wigner matrices.}

\begin{figure}
    \centering
    \includegraphics[width=0.5\linewidth]{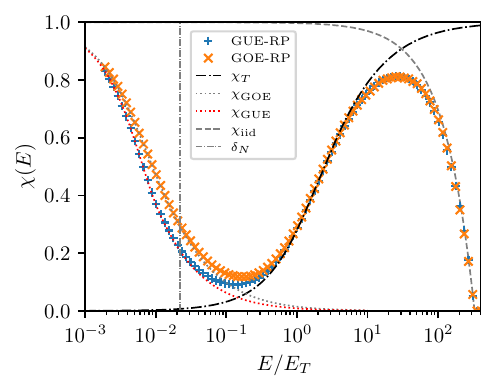}
    \caption{\rev{Comparison between the level compressibility $\chi(E)$ of the GOE-RP and GUE-RP models. 
    Symbols denote data from exact numerical diagonalization of random sample matrices of size $N=30000$, with $\gamma=1.5$, and where the distribution of the diagonal elements is uniform with unit variance, $p_a(a)=1/(2\sqrt{3})$.
    The elements of the matrix $\mathbf{M}$ in \cref{eq:RPdef} belong either to the GOE or GUE, in such a way that $\free_2=1$ in both cases.
    For energies of the scales of the Thouless energy $E_T$~\eqref{eq:thouless}, both curves collapse on the universal prediction $\chi(E) = \chi_T(E/E_T)$ given in \cref{eq:chi_universal}, while for $E\sim \order{1}$, the level compressibility approaches the prediction~\eqref{eq:chi-independent}, corresponding to the case of independent energy levels.
    For energy intervals of the order of the mean level spacing $\delta_N=1/(Np_a(0))$ and below, 
    the replica-based prediction 
    (which relies on the saddle-point approximation) is not expected to hold;
    here the level compressibility approaches instead either the GOE prediction $\chi(E) = \chi_{\rm GOE}(E/\delta_N)$ in Eq.~\eqref{eq:chi_GOE}, or $\chi(E) = \chi_{\rm GUE}(E/\delta_N)$ in \cref{eq:chi_GUE}, respectively.}
    }
    \label{fig:GUE_compressibility}
\end{figure}

\rev{To test these statements, in \cref{fig:GUE_compressibility} we compared two instances of the RP model where the diagonal matrix $\mathbf{A}$ has uniformly distributed entries with unit variance, while the matrix $\mathbf{M}$ is drawn either from the GOE or GUE (the latter with the off-diagonal matrix elements rescaled by a factor $\sqrt{2}$, so that its average spectral density coincides with that of its GOE counterpart). The corresponding level compressibility is measured via exact numerical diagonalization as described in Section~\ref{sec:numerics-free}.
As expected, the curves corresponding to the GOE/GUE cases in \cref{fig:GUE_compressibility} coincide for all energies $E \gg \mathcal O (1/N)$.
By contrast, for
intervals of the order of the mean level spacing $1/(Np_a(0))$ and smaller, 
the saddle-point approximation involved in the replica calculations
is expected to 
break down --- see for instance the discussion in Section~3.6 of Ref.~\cite{Venturelli_2023}.
In this regime, the level compressibility for the GOE case is known instead to reduce to the prediction~\eqref{eq:chi_GOE}, while a similar analytic expression can be obtained for the GUE case by using the results of Ref.~\cite{smith2021full}.
Indeed,
from Eqs.~(31)--(36) therein, one obtains
\begin{equation}
    \label{eq:number_variance_GUE}
    \expval{ I_N[-E,E]^2 } - \expval{I_N[-E,E]}^2 = - \int_{-E}^E  \mathrm{d}x \left ( \int_{-\infty}^{-E} \mathrm{d}y + \int^{\infty}_{E} \mathrm{d}y \right ) C(x,y)
    \; ,
\end{equation}
with
\begin{equation}
    C(x,y) \underset{E \lesssim 1/N}{\approx} - N p_a^2(0) Y_{22} (N p_a(0) |x-y|) \; , 
    \qquad Y_{22}(r) \equiv \frac{\sin^2 (\pi r)}{\pi ^2r^2}
    \; .
\end{equation}
Performing the successive integrals in Eq.~\eqref{eq:number_variance_GUE} and combining with $\expval{I_N[-E,E]} \approx 2 N p_a(0) E$ (see Eq.~\eqref{eq:exp_I_N}), we obtain
\begin{equation}
    \chi(E) = \frac{\expval{ I_N[-E,E]^2 } - \expval{I_N[-E,E]}^2}{\expval{I_N[-E,E]}} \approx \chi_{\rm GUE}(y = Np_a(0) E)
    \; ,
\end{equation} 
where the function $\chi_{\rm GUE}(y)$ is given by
\begin{align}
    \label{eq:chi_GUE}
        \chi_{\rm GUE}(y) =  1 - \frac{2}{\pi}\,\text{Si}(4\pi y) + \frac{\ln(4\pi y) + \gamma_E + 1 - \cos(4\pi y) - \text{Ci}(4\pi y)}{2\pi^2 y} \; .
\end{align} 
Here, ${\rm Ci}(z) = - \int_z^\infty \dd t \, \cos(t)/t $ and ${\rm Si}(z) = \int_0^z \dd t \, \sin(t)/t $ 
    are the cosine-integral and sine-integral functions, respectively, while $\gamma_E$ is the Euler–Mascheroni constant.
}

\section{Derivation of the CGF for energies $E\sim \mathcal{O}(1)$}
\label{app:E-O-1}

In \cref{sec:scaling-function},
we have computed the cumulant generating function of the number of eigenvalues in an interval $[-E,E]=[-x\etanew,x\etanew]$, with $x\sim \order{1}$. This allowed us 
to express the cumulant generating function in a scaling form, \cref{eq:scaling-function-def}, assumed for $E\sim E_T$. In this Appendix, we show that the calculation can be easily generalized to access energies $E\sim \order{1}$ --- similarly to what was done in Ref.~\cite{Venturelli_2023} for the standard GRP model, but with some distinctions which we discuss in \ref{app:comparison}.

\subsection{Saddle-point calculation}
\label{subsec:saddle-point-calc}

Let us go back to the calculation of the cumulant generating function and, in particular, to~\cref{eq:spe-q}. Again, we specialize the calculation to the symmetric interval 
$[\alpha,\beta]=[-E,E]$, but this time with $E\sim \order{1}$, and we decompose the matrix $\underline \Lambda$ in 
Eq.~\eqref{eq:def_lambda_L} as 
\begin{equation}
    \underline \Lambda \equiv E\, \underline{J} - \mathrm i \varepsilon\, 1_{n}
    \; ,
    \label{eq:decomp-Lambda-1}
\end{equation}
with $\underline{J}$ given in \cref{eq:def_lambda_J}. We then replace $q_k=q_k\z+\order{\etanew}$ into the saddle-point equation~\eqref{eq:spe-q} and discard all terms of $\order{\etanew}$, finding 
\begin{equation}
    q_k\z = - \frac{2}{\mathcal Z_\varphi\z} \int \dd{a} \, p_a(a) \int \dd{\vec y} \, y_k^2 \, \exp[-\frac{1}{2}\vec y^{\; T} \left( \varepsilon\, 1_{n} - \mathrm i a \underline L + \mathrm i E\, \underline{J} \right)\,\vec y] 
    \; ,
    \label{eq:spe-q-1}
\end{equation}
where
\begin{align}
    \label{eq:Z0-1}
    &\mathcal Z_\varphi\z = \int \dd{a} \, p_a(a) \int \dd{\vec y} \, \exp[-\frac{1}{2}\vec y^{\; T} \left( \varepsilon\, 1_{n} - \mathrm i a \underline L + \mathrm i E\, \underline{J} \right)\,\vec y] \\
    &=(2\pi)^\frac{n}{2} \int \dd{a} \, p_a(a) \left\lbrace
    \left[ \varepsilon-\mathrm i (a+E)  \right]
    \left[  \varepsilon+\mathrm i (a-E)  \right]
    \right\rbrace^{-\frac{n_+}{2}} \left\lbrace
    \left[ \varepsilon-\mathrm i (a-E) \right]
    \left[\varepsilon+\mathrm i (a+E) \right]
    \right\rbrace^{-\frac{n_+}{2}}\n\\
    &\xrightarrow[n_\pm \to \pm \mathrm i s/\pi]{}  \int \dd{a} p_a(a) \exp{\frac{\mathrm i s}{2\pi} \ln\frac{\left[ \varepsilon-\mathrm i (a-E)  \right]
    \left[\varepsilon+\mathrm i (a+E)   \right]}
    {\left[ \varepsilon-\mathrm i (a+E)  \right]
    \left[ \varepsilon+\mathrm i (a-E) \right]}}
    \equiv \int \dd{a} p_a(a) \mathcal Z_\varepsilon(a)
    \; . \n
\end{align}
We note for future convenience that $\mathcal Z_\varepsilon^*(a)
=\mathcal Z_\varepsilon(a)$; moreover,
$\mathcal Z_\varepsilon(a)\xrightarrow[]{\varepsilon\to 0^+} 1  $, 
and hence $\mathcal Z_\varphi\z \xrightarrow[]{\varepsilon\to 0^+} 1$. 
We can now simplify \cref{eq:spe-q-1} 
using
\begin{align}
    &\int \dd{a} \, p_a(a) \int \dd{\vec y} \, y_k^2 \, 
    \exp[-\frac{1}{2}\vec y^{\; T} \left( \varepsilon\, 1_{n} - \mathrm i a \underline L + \mathrm i E\, \underline{J}\right)\,\vec y] \n\\
    &\xrightarrow[n_\pm \to \pm \mathrm i s/\pi]{} \int \dd{a} \frac{p_a(a) \mathcal Z_\varepsilon(a)}{\varepsilon-\mathrm i a \ell_{kk} + \mathrm i E j_{kk}}
    \; ,
\end{align}
where $j_{kk}$ are the matrix elements of $\underline{J}$ in \cref{eq:def_lambda_J}.
We thus find from~\cref{eq:spe-q-1}
\begin{equation}
    q_k\z 
    = -\frac{2}{\int \dd{a} p_a(a) \mathcal Z_\varepsilon(a)} \int \dd{a} \frac{p_a(a) \mathcal Z_\varepsilon(a)}{\varepsilon-\mathrm i a \ell_{kk} + \mathrm i E j_{kk}} 
    \; .
\end{equation}
To make progress, we now assume that $p_a(-a)=p_a(a)$, so that in particular
$\mathcal Z_\varepsilon(-a)
=\mathcal Z_\varepsilon(a)$,
and we plug in explicitly the four possible values of $\ell_{kk}$ and $j_{kk}$. It is then evident that $q\z_2 = q_1\z$ and $q_3\z=q_4\z=[q_1\z]^*$. In particular, 
\begin{align}
    q_1\z &= -\frac{2}{\int \dd{a} p_a(a) \mathcal Z_\varepsilon(a)} \int \dd{a} \frac{p_a(a) \mathcal Z_\varepsilon(a)}{\varepsilon-\mathrm i a  - \mathrm i E } 
    = \frac{2\mathrm i}{\int \dd{a} p_a(a) \mathcal Z_\varepsilon(a)} \int \dd{a} \frac{p_a(a) \mathcal Z_\varepsilon(a)}{a-(E+\mathrm i \varepsilon) } \n\\
    &\xrightarrow[\varepsilon\to 0^+]{} -2\mathrm i \mathcal{G}_a(E+\mathrm i 0^+) \equiv -\mathfrak q
    \; ,
    \label{eq:def-qfrak}
\end{align}
where in the second line we recognized the resolvent 
\begin{equation}
    \mathcal{G}_a(z) = \int \dd{a} \; \frac{p_a(a)}{z-a}
    \label{eq:resolvent}
\end{equation}
associated to the distribution $p_a(a)$. We stress that $\mathfrak q$ (defined by \cref{eq:def-qfrak})
is actually a complex number (i.e.~not purely imaginary), and that the $+\mathrm i 0^+$ prescription inside the argument of the resolvent is necessary both to ensure its convergence and to specify the correct branch.

We are now in a position to compute the action in Eq.~\eqref{eq:action-Q}. The first term gives
\begin{equation}
    \frac{\etanew \cum_2}{16} \Tr (\underline Q \underline L)^2  
    = 
    \frac{\etanew \cum_2}{16} \sum_{ik}Q_{ik}^2 \ell_{ii}\ell_{kk} 
    \xrightarrow[n_\pm \to \pm \mathrm i s/\pi]{\varepsilon\to 0^+} 
    -\frac{s\etanew \cum_2}{4\pi}\Im{\mathfrak q^2}+\mathcal{O}(\etanew^2)
    \; ,
    \label{eq:trace}
\end{equation}
where we used the optimized solution $Q_{ik} = q\z_k \delta_{ik} + \order{\etanew}$.
The problem then reduces to computing $\mathcal Z_\varphi\o$
as in \cref{eq:S-lnZ},
where $Z_\varphi\o$ can be found by inserting the optimized solution for $Q_{ik}$ into Eq.~\eqref{eq:Zphi}. 
Upon computing the Gaussian integral over $\vec y$, we find 
\begin{align}
    \mathcal Z_\varphi\o&=  \int \dd{a} p_a(a) (2\pi)^\frac{n}{2} \left\lbrace
    \left[ \varepsilon -\etanew \cum_2 q_1\z/2 -\mathrm i (a+E)  \right]
    \left[  \varepsilon -\etanew \cum_2 q_2\z/2 +\mathrm i (a-E)  \right]
    \right\rbrace^{-\frac{n_+}{2}} \n\\
    & \qquad \qquad\qquad \quad \times \left\lbrace
    \left[ \varepsilon -\etanew \cum_2 q_3\z/2-\mathrm i (a-E) \right]
    \left[\varepsilon -\etanew \cum_2 q_4\z /2+\mathrm i (a+E) \right]
    \right\rbrace^{-\frac{n_-}{2}}\n\\
    &\xrightarrow[n_\pm \to \pm \mathrm i s/\pi]{\varepsilon \to 0^+}  \int \dd{a} p_a(a) \exp{\frac{\mathrm i s}{2\pi} \ln\frac{
    \left[ \etanew\cum_2 \mathfrak q^*-2\mathrm i (a-E)   \right]
    \left[  \etanew\cum_2 \mathfrak q^*+2\mathrm i (a+E)  \right]}
    {\left[ \etanew\cum_2 \mathfrak q-2\mathrm i (a+E)   \right]
    \left[  \etanew\cum_2 \mathfrak q+2\mathrm i (a-E)  \right]}}
    \; .
    \n
    \\
    \label{eq:Zphi-step-1}
\end{align}
Next, 
we introduce the parameters $r$ and $\theta$ as in \cref{eq:r-theta},
so as to rewrite \cref{eq:Zphi-step-1} as 
\begin{equation}
    \mathcal Z_\varphi\o \xrightarrow[n_\pm \to \pm \mathrm i s/\pi]{\varepsilon \to 0^+} \int_{-\infty}^\infty \dd{a} p_a(a) \exp[ -\frac{s}{\pi} \arctan \left( \frac{ \sin 2\theta}{a^2r^2 + \cos 2\theta} \right) ]
    \; ,
\end{equation}
where we used the identity $2\mathrm i \arctan(z) = \ln[(1-\mathrm i z)/(1+\mathrm i z)]$. 
Using \cref{eq:F-S,eq:action-Q}, at leading order for small $\etanew$, we then find 
the expression of $\mathcal F_{[-E,E]} (s)$ reported in \cref{eq:CGF-E-O-1}.

\subsection{Comparison with Ref.~\cite{Venturelli_2023}}
\label{app:comparison}

The result obtained in~\ref{subsec:saddle-point-calc},
i.e.~the cumulant generating function~\eqref{eq:CGF-E-O-1},
should be compared to the one previously obtained by some of us in Section~3.3 of Ref.~\cite{Venturelli_2023}. A few comments are in order.

First, we note that the result~\eqref{eq:CGF-E-O-1} obtained here is in closed form, whereas the one in~\cite{Venturelli_2023} required to further solve a self-consistency equation. This is because here we decided to retain only the leading-order contribution in powers of $\etanew$ throughout the entire calculation, which (as it turns out) allows us to close the self-consistency equations.
The order of magnitude of the discarded terms is thus different: here it is of $\order{\etanew^2}= \order{N^{2(1-\gamma)}}$, whereas in Ref.~\cite{Venturelli_2023} the error was determined by the Gaussian fluctuations around the saddle point, which were estimated to be of $\order{\etanew/N}= \order{N^{-\gamma}}$ (note that $\eta \propto \etanew$ in the notation of Ref.~\cite{Venturelli_2023}). 
Since $N^{2(1-\gamma)}>N^{-\gamma}$ for $\gamma<2$, the estimate given here is more conservative.

However, we reckon that this new estimate of the order of magnitude of the 
error
is the correct one to be retained. Indeed, we realized that Eq.~(87) in Ref.~\cite{Venturelli_2023} inadvertently involved an approximation, and is in fact only valid at leading order for small $\etanew$. The same is true for an analogous step in the calculation for the Wishart--RP model in Ref.~\cite{Delapalme2026}, see Eqs.~(85)~and~(B.39) therein. This does not change in any way the results reported in Refs.~\cite{Venturelli_2023,Delapalme2026},
but merely the estimate of the error in the analytical calculation (which is in any case generally negligible \rev{already for moderately large $N$}, as was amply checked via numerical diagonalization).

\section{Full counting statistics for independent energy levels}
\label{app:FCS-independent}

If the energy levels are independent, then the full counting statistics, together with its finite-size corrections, can be computed analytically, as we show in this Appendix.

Consider $N$ independent random variables $x_1,\ldots,x_N$, each
drawn uniformly from the box $[-W/2,\,W/2]$.  Let $I_N$ denote the
number of them that fall inside an interval of half-width $\delta/2$
centred at the origin, i.e.\ in $[-\delta/2,\,\delta/2]$. The probability of a single variable landing in the interval is $p = \frac{\delta}{W}$, so $I_N$ is exactly Binomial$(N,p)$:
\begin{equation}
  P(I_N) = \binom{N}{I_N} p^{I_N}(1-p)^{N-I_N} \, .
\end{equation}
To make contact with the analysis of Fig.~\ref{fig:counting:rare}, in which the energy window is set to be of the order of the Thouless energy $E_T \propto N^{1-\gamma}$ (see Eq.~\eqref{eq:thouless}) for $\gamma = 1.5$, we choose
\begin{equation}
  \delta = \frac{a}{\sqrt{N}} \, , \qquad a = \mathcal O(1) \, ,
\end{equation}
so that $p = a/(W\sqrt{N}) \to 0$ as $N\to\infty$.  The mean number
of elements in the interval $[-\delta/2,\,\delta/2]$ is
\begin{equation}
  \langle I_N \rangle \equiv \lambda = Np = \frac{a\sqrt{N}}{W} \;\sim\; \sqrt{N } \, ,
\end{equation}
which grows with $N$, though more slowly than $N$ itself. We introduce the scaled variable
\begin{equation}
  k \equiv \frac{I_N W}{\sqrt{N}} \, , \qquad k = \mathcal O(1) \, ,
\end{equation}
so that $I_N = k\sqrt{N}/W = k \lambda / a$. Because $p\to 0$ while $\lambda\to\infty$, the Binomial converges
to a Poisson distribution with mean $\lambda$:
\begin{equation}\label{eq:poisson}
  P\!\left(I_N = \frac{k\sqrt{N}}{W}\right)
  \;\approx\;
  \frac{e^{-\lambda}\,\lambda^{I_N}}{I_N!} \, .
\end{equation}

We now seek a large-deviation representation of the form of Eq.~\eqref{eq:def_ldf},
\begin{equation}\label{eq:ldf}
  P\!\left(I_N = \frac{k\sqrt{N}}{W}\right)
  = \exp\!\left(-\frac{\sqrt{N}}{W}\,\Phi(k)\right) \, ,
\end{equation}
where $\Phi(k) \geq 0$ is the rate function with $\Phi(a)=0$.
Taking the logarithm of Eq.~\eqref{eq:poisson} and introducing
the shorthand $s \equiv \sqrt{N}/W$, so that $I_N = ks$ and
$\lambda = as$, we obtain
\begin{equation}
  -\ln P
  \;=\;
  \lambda - I_N \ln\lambda + \ln(I_N!)
  \;=\;
  as - ks\ln(as) + \ln\!\bigl[(ks)!\bigr].
\end{equation}
Using the full Stirling expansion,
$\ln(n!) = n\ln n - n + \tfrac{1}{2}\ln(2\pi n) + \mathcal O(1/n)$,
the last term reads
\begin{equation}
  \ln\!\bigl[(ks)!\bigr]
  = ks\ln(ks) - ks + \tfrac{1}{2}\ln(2\pi ks) +\mathcal O\!\left(\frac{1}{ks}\right).
\end{equation}
Assembling all contributions we find
\begin{align}
  -\ln P
  &= as - ks\ln(as) + ks\ln(ks) - ks
     + \tfrac{1}{2}\ln(2\pi ks) + \cdots \notag\\
  &= s\!\underbrace{\Bigl[k\ln\frac{k}{a} - k + a\Bigr]}_{\Phi(k)}
     + \tfrac{1}{2}\ln(2\pi ks) + \cdots\,.
\end{align}
Identifying $s = \sqrt{N}/W$, the full asymptotic expression is thus
\begin{equation}\label{eq:full}
  P\!\left(I = \frac{k\sqrt{N}}{W}\right)
  \;\approx\;
  \frac{1}{\sqrt{2\pi k\sqrt{N}/W}}\,
  \exp\!\left(-\frac{\sqrt{N}}{W}\,\Phi(k)\right),
\end{equation}
where the prefactor carries the leading sub-exponential correction. In practice, to retain all Stirling corrections at finite $N$, we
can write
\begin{eqnarray}\label{eq:psiN}
  \Phi_N(k)
  &\equiv& 
  -\frac{W}{\sqrt{N}}\ln P\!\left(\frac{k\sqrt{N}}{W}\right)
  \nonumber\\
  &=& 
  \frac{W}{\sqrt{N}}\Bigl[
      \frac{a\sqrt{N}}{W}
      - \frac{k\sqrt{N}}{W}\ln\!\frac{a\sqrt{N}}{W}
      + \ln\Gamma\!\!\left(\frac{k\sqrt{N}}{W}+1\right)
    \Bigr].
\end{eqnarray}
As $N\to\infty$ (with $k$ and $a$ fixed), only the term linear in
$s = \sqrt{N}/W$ survives, and one recovers the Poisson rate
function in the variable $k$:
\begin{equation}\label{eq:psiinf}
  \Phi(k) \equiv \lim_{N\to\infty} \Phi_N(k)
         = k\ln\frac{k}{a} - k + a.
\end{equation}
Note that $\Phi(k) \to a$ for $k \to 0$.

\begin{figure}
    \centering
    \includegraphics[width=0.54\linewidth]{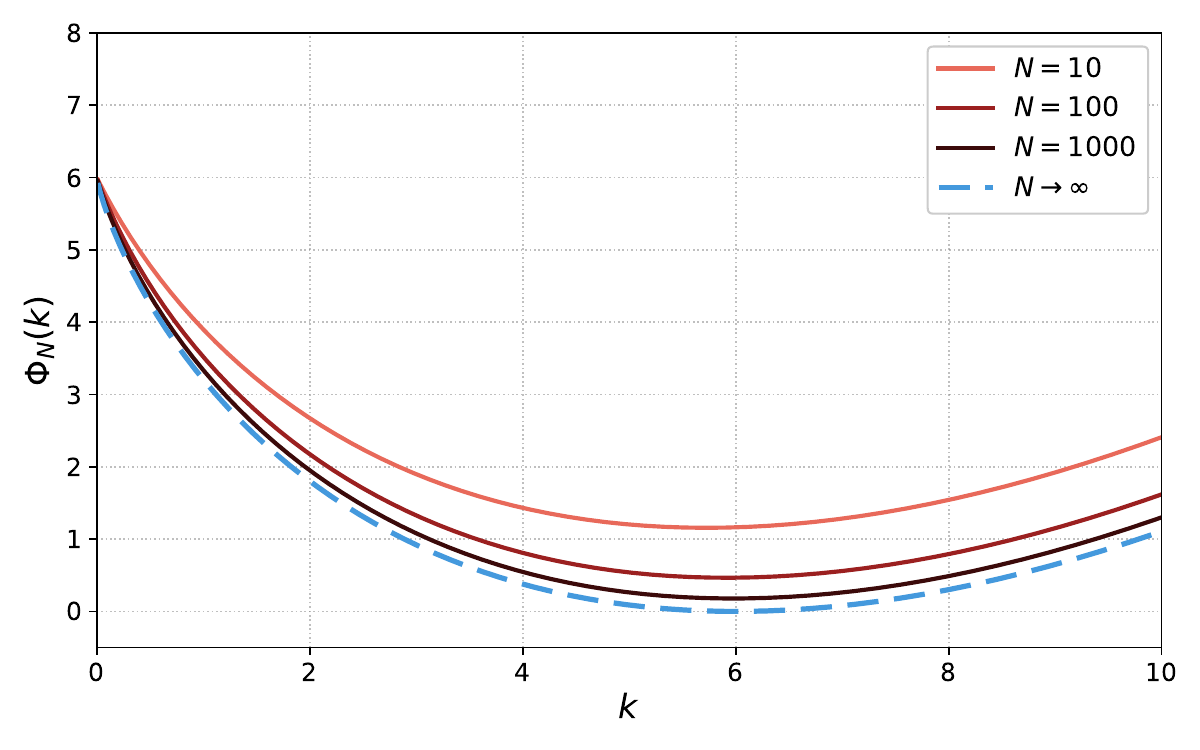}
    \caption{Rate function $\Phi_N(k)$ from Eq.~\eqref{eq:psiN}, for
  $W=\sqrt{3}$, $a=6$, and $N=10,100,1000$ (solid, light to dark red),
  together with the $N\to\infty$ limit $\Phi(k)$ from
  Eq.~\eqref{eq:psiinf} (blue dashed curve).
  All curves attain their minimum at $k=a=6$.
  The finite-$N$ corrections are purely additive and decrease as
  $W/(2\sqrt{N}) \, \ln(2\pi k\sqrt{N}/W)$, becoming negligible
  for large $N$. }
    \label{fig:psi}
\end{figure}

From this expression for $\Phi(k)$, 
one easily identifies its minimum: $\Phi'(k) = \ln(k/a) = 0 \Rightarrow k^* = a$,
        confirming that the most probable value is $I_N = a\sqrt{N}/W = \lambda$,
        as expected.
Expanding around $k^*=a$, one also obtains the Gaussian fluctuations:
        \begin{equation}
          \Phi(k) \approx \frac{(k-a)^2}{2a}
          + \mathcal O\!\left((k-a)^3\right),
        \end{equation}
        since $\Phi''(k)=1/k$ and $\Phi''(a)=1/a$.  This form is consistent
        with the Poisson distribution property ``variance $=$ mean $= \lambda = a\sqrt{N}/W$'' (therefore corresponding to a level compressibility equal to one).

Figure~\ref{fig:psi} shows $\Phi_N(k)$. To make contact with the analysis of Section~\ref{sec:counting:rare}, we chose the same parameters as in Fig.~\ref{fig:counting:rare}, i.e.~$W=\sqrt{3}$, $a=6$. Finite-$N$ rate functions are computed from
Eq.~\eqref{eq:psiN} for
$N=10,\,100,\,1000$, and the asymptotic limit $\Phi(k)$
from Eq.~\eqref{eq:psiinf}.  All curves share the minimum
$\Phi_N(a)=0$ at $k=a=6$, and converge to the asymptotic form
as $N$ increases.  The finite-$N$ corrections raise the curve
uniformly, reflecting the logarithmic Stirling prefactor in
Eq.~\eqref{eq:full}.

\section*{References}
\bibliographystyle{iopart-num}
\bibliography{references} 

\end{document}